\documentclass[11pt]{article}

\usepackage[english]{babel}
\usepackage[utf8]{inputenc}
\usepackage[usenames,dvipsnames]{color}
\usepackage{adjustbox}

\usepackage[a4paper,left=2.65cm,right=2.65cm,top=2.5cm,bottom=3cm,bindingoffset=0mm]{geometry}

\usepackage{cite}

\usepackage{amsthm}
\usepackage{amssymb}

\usepackage{extarrows}

\usepackage[normalem]{ulem}
\usepackage{cancel}
\usepackage[arrow, matrix, curve]{xy}
\usepackage{stmaryrd}
\usepackage{eurosym}
\usepackage{graphicx}
\usepackage[bookmarksopen,plainpages=false,pdfpagelabels]{hyperref}
\usepackage{tikz}
\usepackage{tikz,fullpage}
\usetikzlibrary{arrows,%
                petri,%
                topaths}%
\usepackage{tkz-berge}
\usepackage{subfig}

\usepackage[onehalfspacing]{setspace}

\begin{document}

\vspace*{-1.5cm}
\begin{flushright}
  {\small
  LMU-ASC 11/20\\
  MPP-2020-34
  }
\end{flushright}

\vspace{1.75cm}

\begin{center}
{\LARGE
Detecting Symmetries with Neural Networks}
\end{center}

\vspace{0.4cm}

\begin{center}
  Sven Krippendorf$^1$, Marc Syvaeri$^{1,2}$ 
\end{center}

\vspace{0.3cm}

\begin{center} 
\textit{$^{1}$\hspace{1pt} Arnold Sommerfeld Center for Theoretical Physics\\[1pt]
Ludwig-Maximilians-Universit\"at \\[1pt]
Theresienstra\ss e 37 \\[1pt]
80333 M\"unchen, Germany}
\\[1em]
\textit{$^{2}$\hspace{1pt} Max-Planck-Institut f\"ur Physik\\[1pt]
F\"ohringer Ring 6 \\[1pt]
80805   M\"unchen, Germany}
\end{center} 

\vspace{0.8cm}

%%%%%%%%%%%%%%%%%%%%%%%%%%%%%%%%%%%%%%%%%%%%%%%
%%%%%%%%%%%%%%%%%%%%%%%%%%%%%%%%%%%%%%%%%%%%%%%
%%%%%%%%%%%%%%%%%%%%%%%%%%%%%%%%%%%%%%%%%%%%%%%
%%%%%%%%%%%%%%%%%%%%%%%%%%%%%%%%%%%%%%%%%%%%%%%

\begin{abstract}
\noindent
Identifying symmetries in data sets is generally difficult, but knowledge about them is crucial for efficient data handling. Here we present a method how neural networks can be used to identify symmetries. We make extensive use of the structure in the embedding layer of the neural network which allows us to identify whether a symmetry is present and to identify orbits of the symmetry in the input. To determine which continuous or discrete symmetry group is present we analyse the invariant orbits in the input. We present examples based on rotation groups $SO(n)$ and the unitary group $SU(2).$ Further we find that this method is useful for the classification of complete intersection Calabi-Yau manifolds where it is crucial to identify discrete symmetries on the input space. For this example we present a novel data representation in terms of graphs.
\end{abstract}

\newpage
\tableofcontents

\section{Introduction}
\label{sec:introduction}

One ubiquitous feature in nature is the presence of symmetries, ranging from the ultra-small captured by the symmetries underlying the Standard Model of Particle Physics to the isotropy and homogeneity of our Universe on cosmological scales; and in every day life when one wants to identify objects in a picture with a neural network. The question we pursue in this paper is: {\it Can we use neural networks to detect symmetries in an underlying data product?}

We present a method which is suitable for data questions where we have samples of a function of the input variables $f(x_{\rm input}).$ This situation is present in supervised learning. The presence of a symmetry is simply the statement that inputs which are transformed under some symmetry transformation $x_{\rm input}\to S(x_{\rm input})$ lead to the same output $f(S(x_{\rm input}))=f(x_{\rm input}).$

The key idea which we utilise to find symmetries, is the fact that objects which are invariant under symmetries are clustered together in the embedding space (i.e.~the second to last layer in our neural networks). As a first step, this reveals the presence of symmetries. Effectively, this is rather similar to word embeddings found in word2vec~\cite{mikolov2013efficient}, which has also been utilised to identify similarities between chemical elements~\cite{ZhouE6411}.
By analysing the relation of the points in the input space we are then able to identify the nature of the symmetry, i.e.~we determine the generators of the symmetries.

We test this method on artificial datasets with an underlying rotational group $SO(2)$ and $SO(3),$ and show how we can identify a unitary group (here: $SU(2)$) and distinguish it from larger symmetry groups (here: $SO(4)$). To show the applicability of the identification of generators in higher dimensional datasets (e.g.~images), we discuss how we can identify $SO(2)$ in the context of rotated MNIST data.

We use this method in the context of the classification of consistent vacua in string theory. Finding distinct ways to obtain string vacua is a crucial step in improving our understanding of string theory as a theory of quantum gravity. One aspect is the classification of consistent string backgrounds, in particular Calabi-Yau manifolds (CYs). To obtain a classification one needs to remove redundancies arising from multiple representations of the same manifold. We apply our method to the case of complete intersection Calabi-Yau manifolds (CICYs). Utilising a novel representation in terms of graph networks, we perform the supervised classification task for two topological invariants, the Hodge numbers $h^{1,1}$ and $h^{1,2}.$ When analysing the embedding layer, we are able to re-identify the known identities in the dataset.

The rest of the paper is organised as follows. In Section~\ref{sec:symmetries} we describe how symmetries  can be found in the embedding layer. We then examine the orbits in the input layer to identify the underlying symmetry in Section~\ref{sec:generators}, before presenting our conclusions.

\section{Finding Symmetries}
\label{sec:symmetries}
In this section we present a method of how to identify previously `unknown’ symmetries in  a dataset by examining the clustering behaviour in the embedding layer.  We study this method on two types of examples -- continuous and discrete symmetries.

In the first part, we discuss two examples based on real and complex-valued functions. For this we take the Mexican hat potential in two dimensions which features an $SO(2)$-symmetry, and an $SU(2)$ invariant superpotential (holomorphic function). The procedure to find symmetries is as follows: Within these potentials, we define classes which are defined by a respective value of the potential. This enables us to construct a classification problem.\footnote{In our experiments, we find that regression does not expose symmetries in the same way.}  We train our network to address this classification task and examine the representation in the embedding layer. This reveals that the representation distinguishes between points connected via the symmetry and points not connected but still in the same class. Coarsely speaking, the network clusters symmetry invariant points and there is a gap in the embedding layer to the other points in the class.

In the second part, we study discrete symmetries in the context of classification of CICYs in three dimensions. We take multiple representatives of each manifold, and train the network to classify some topological invariants, the Hodge numbers $h^{1,1}$ and $h^{1,2}.$
Again, by analysing the structure of the embedding layer, we are able to identify finer grained classes compared to the trained classes. These finer grained classes are comprised of different representatives of the respective CICY manifold. The neural network must use other quantities which it is not trained on.

Depending on the dimension of the embedding space, we use a  dimensional reduction with TSNE~\cite{tsne} to be able to plot the data points and to visualise its structures.

This identification of a symmetry in the dataset is then  used in a second step to construct the generators associated with this symmetry. This is discussed in Section~\ref{sec:generators} and this step allows us to identify the underlying symmetry.

\subsection{Continuous Symmetries}
\subsubsection*{Mexican-Hat-Potential}
\begin{figure}%
\centering
\raisebox{0.1\height}{\includegraphics[width=0.45\textwidth]{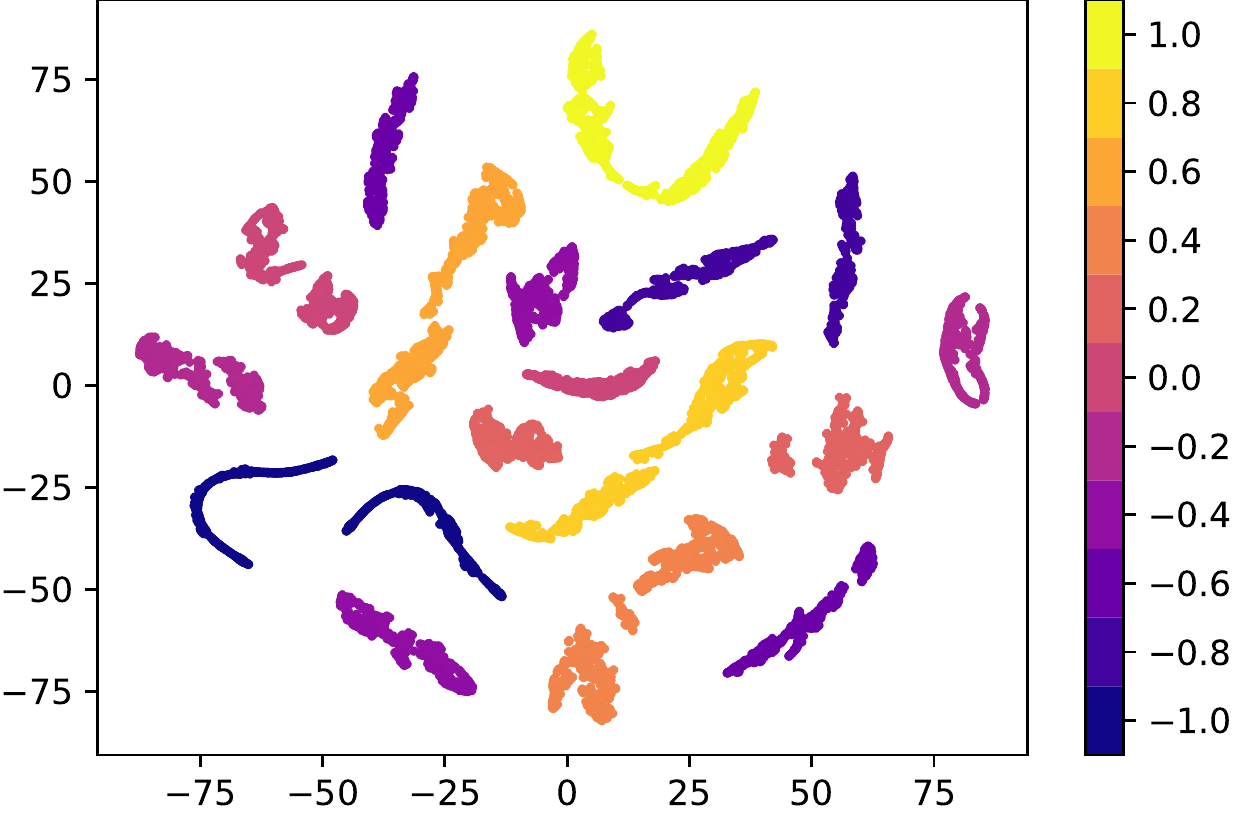}}
\qquad
\includegraphics[width=0.48\textwidth]{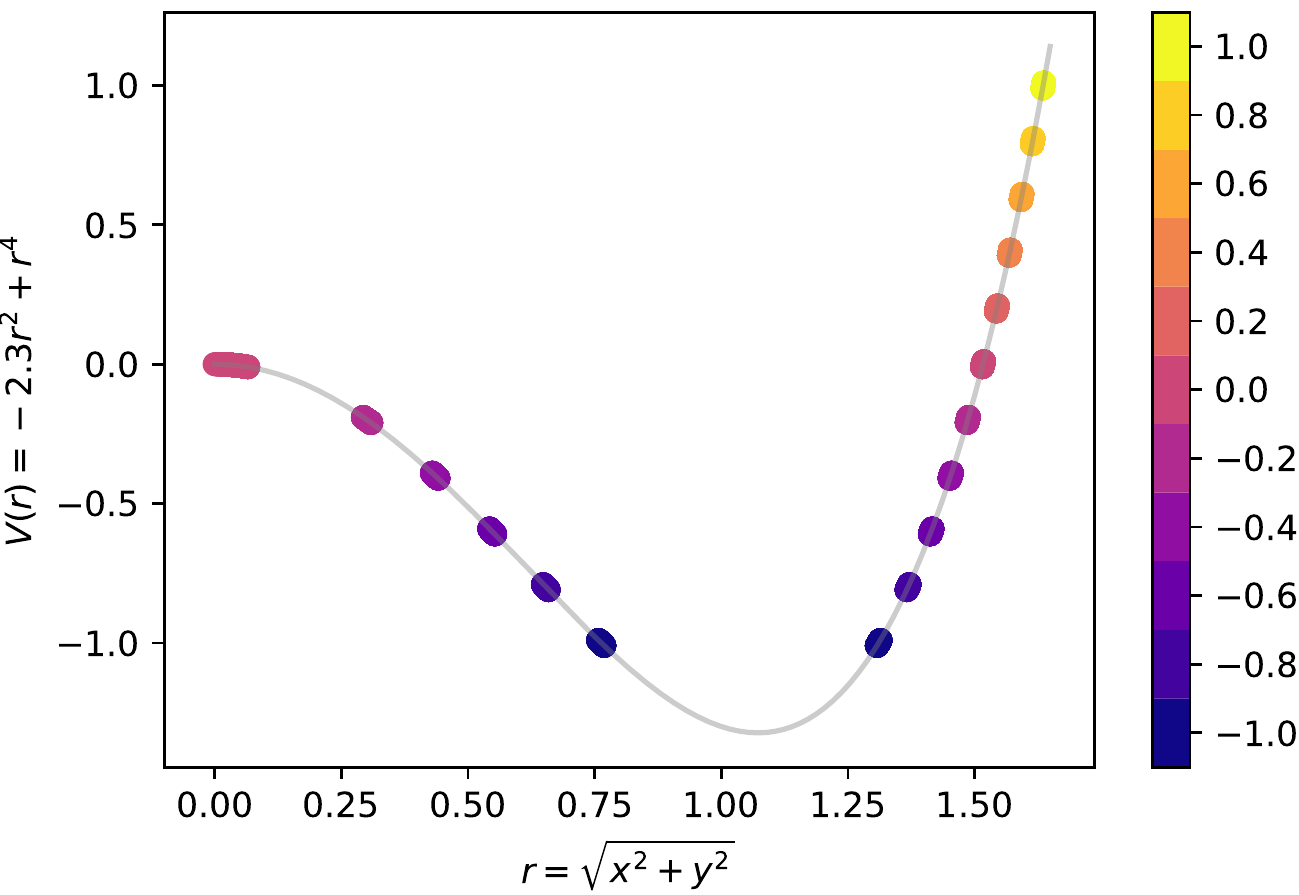}
%\vspace{-0.5cm}
\caption{{\bf Left:} This shows the TSNE-representation (perplexity of $50$) of the embedding layer. Each colour represents one class. For several classes, we can directly see two distinct point clouds. {\bf Right:} This shows the plot of the Mexican hat potential where we highlight the classes using the same color coding as on the left panel. Here, we can directly match points with multiple clusters and disconnected TSNE components.}\label{fig:tsneso2}
\end{figure}

We start with a two dimensional function with an underlying $SO(2)$-symmetry:
\begin{equation}
    V(x,y)=-a\cdot(x^2+y^2)+(x^2+y^2)^2 = -a\cdot r^2+r^4\, ,
\label{eq:potentialso2}
\end{equation}
where we use $a=2.3$ for our numerical experiments.
Here, two types of points appear: Points with the same value for the potential~\eqref{eq:potentialso2} which are related by a symmetry transformation and points which are not related by a symmetry. Examples of such points can be found in the plot of the potential shown in the right panel of  Figure~\ref{fig:tsneso2}.

We formulate our classification problem as follows: we define 11 classes for the function where the values of these classes are as follows:
\begin{equation}
    \begin{split}
        \left[\frac{k}{5}-10^{-3},\frac{k}{5}+10^{-3}\right] \qquad k=-5,...,5\,.
    \end{split}
\end{equation}
Then we sample points by randomly picking values for $x$ and $y$, and checking whether they belong to one of the classes. For training, we use balanced training sets with $\sim 1000$ representatives per class.
We train a simple network consisting of $7$ dense layers with $80$ hidden units with ReLu-activation and a final layer with $11$ dimensional softmax output activation.\footnote{We use tensorflow with keras backend. For PCA and TSNE implementations we use~\cite{scikit-learn}.} We use categorical crossentropy with Adam optimiser.\footnote{Our results do not require large hyperparameter tuning.}
We train our network on this classification task to a reasonable training accuracy (above 95 percent).\footnote{We only define a training dataset because we are only interested in correctly classified data points. At this stage there is no necessity to construct a test or validation set.} We then visualise the representation on the embedding layer by applying TSNE on this $80$-dimensional data set which can be found in Figure~\ref{fig:tsneso2}. 

 Looking at a specific class, one can directly see that the separating property is the norm of the point. To be precise, points bigger than the norm of the minimum of the potential at $r=\sqrt{{a}/{2}}$ are separated by points with smaller norm. In Figure~\ref{fig:tsneso2} we can identify for multiple of these classes that they clearly split in two regions whereas for classes with elements from only `one' radius they are not split.

\subsubsection*{Superpotential}
We now demonstrate the method on an example with an $SU(2)$-symmetry. To do this we examine the following complex valued function
\begin{equation}
\begin{split}
    W(x,y)=(x_1 y_2-x_2 y_1)+\frac{1}{2}(x_1 y_2-x_2 y_1)^2\,,
\end{split}
\end{equation}
where $x=(x_1,x_2),y=(y_1,y_2) \in \mathbb{C}^2$ and transform in the fundamental and anti-fundamental representation of $SU(2)$ respectively. Such holomorphic functions appear for instance in supersymmetric field theories and are referred to as superpotentials. Here we are interested in finding the symmetries in this superpotential. In addition to the $SU(2)-$symmetry, this superpotential has two independent scaling symmetries:
\begin{equation}
\begin{split}
    x_1 &\rightarrow a\,x_1\\
    y_2 &\rightarrow \frac{1}{a}\,y_2
\end{split}
\quad\qquad\qquad\quad
  \begin{split}
            x_2 &\rightarrow b\,x_2\\
    y_1 &\rightarrow \frac{1}{b}\,y_1 \,,
  \end{split}
\end{equation}
where $a,b \neq 0$. However, we check that orbits of these symmetries are not present in our datasets.

\begin{figure}
\begin{center}
\includegraphics[width=0.45\textwidth]{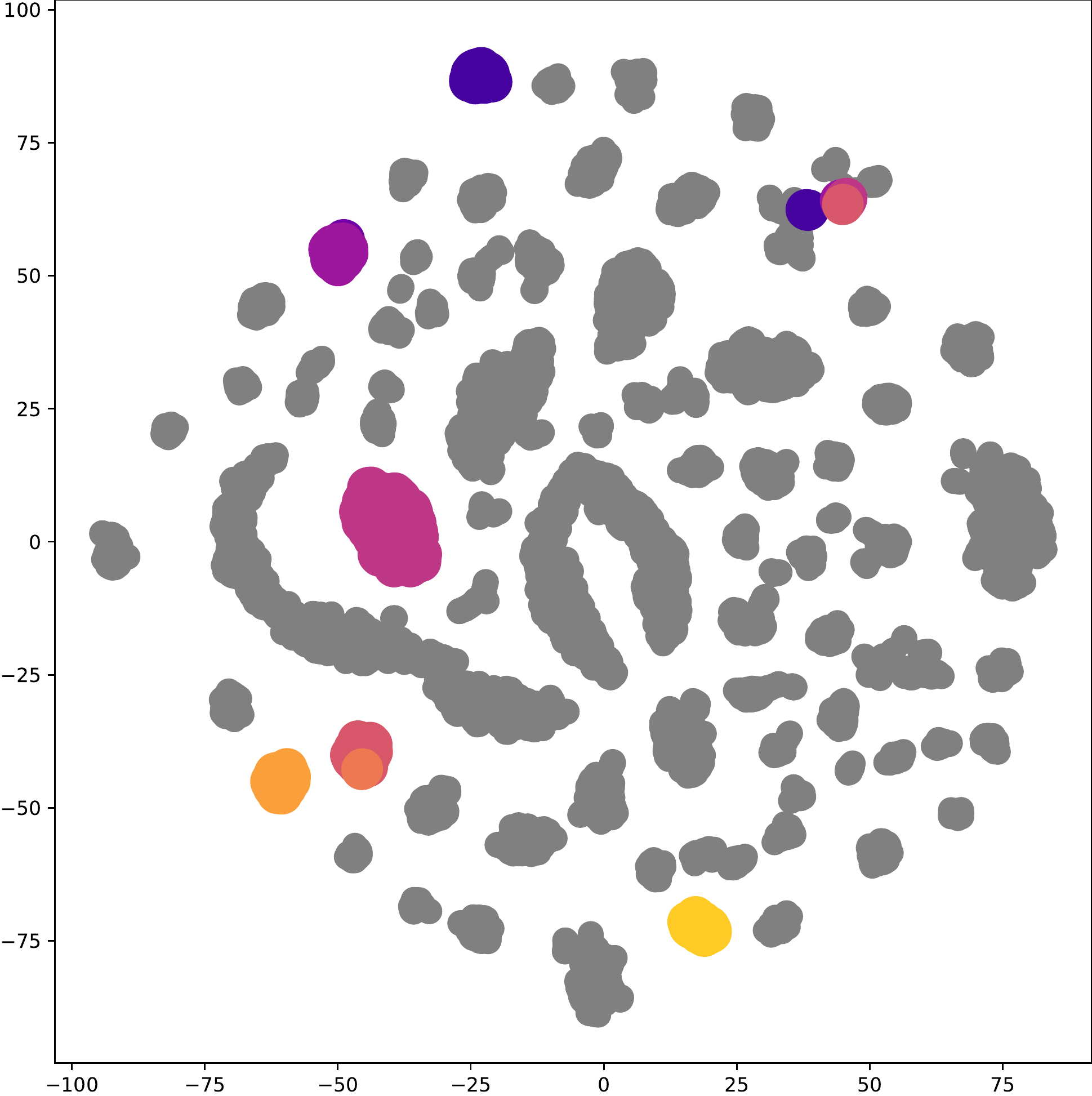}\qquad\includegraphics[width=0.45\textwidth]{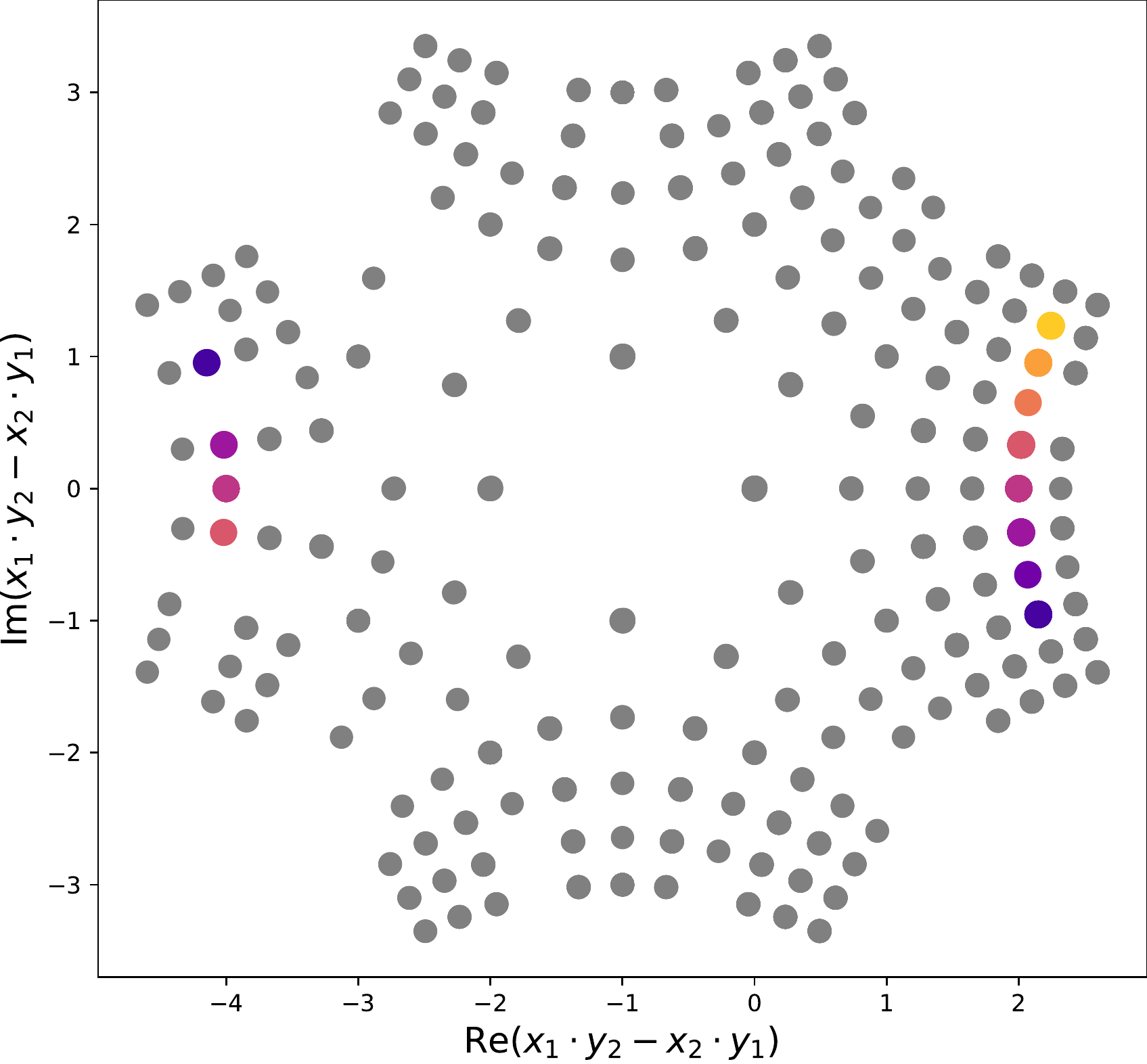}
\end{center}
\caption{{\bf Left:} This is a TSNE-projection of the $60$-dimensional embedding space (perplexity $40$). The coloured dots mark the same classes as highlighted on the right hand side. Gray dots denote the other points in the embedding.
{\bf Right:} $SU(2)$ invariant quantity $x_1\cdot y_2-x_2\cdot y_1.$ Most classes have two distinct representatives but some only have one. For instance, the yellow and light orange class have a single $SU(2)$ invariant. In the embedding layer there are no distinct clusters for these points unlike for the other points.}

\label{fig:tsnesu2}
\end{figure}

Proceeding as before, we firstly sample points for the superpotential and categorise them regarding their outputs. We have one classification with 11 class labels for the real part and one classification for the imaginary part. We choose the following numerical ranges, which are symmetric around zero:
\begin{equation}
\begin{split}
        \left[k- 10^{-2},k+ 10^{-2}\right] \qquad k=-5,...,5\,.    
\end{split}
\end{equation}
With this classification we cover the entire output range in the open subset ${\rm Re}(z),{\rm Im}(z)\in(-5.,5.).$ Again, we sample the points by randomly picking values for $x$ and $y$, and checking whether their real and imaginary part both belong to one of these classes. As in the previous case, we trained a simple network consisting of $7$ dense layers with $60$ neurons and ReLu-activation, followed by two $11$-dimensional dense layers with softmax activation.   
As before, we use categorical crossentropy for each of these output layers with an Adam optimiser. For training we used a balanced set with $\sim 1000$ representatives per class and we terminated training at an accuracy of slightly above $95$ percent. Again, we visualise the structure of the $60$-dimensional embedding layer by applying TSNE and show the resulting two dimensional space in Figure~\ref{fig:tsnesu2}.

In this projection, it is tedious to find different regions as a consequence of having $121$ different classes. We highlight some examples of the separation in the point clouds in Figure~\ref{fig:tsnesu2} with one and two distinct $SU(2)$ representatives respectively. This can be seen by computing the invariant quantity of $SU(2)$ $\epsilon_{ij}x^i y^j$ (where $\epsilon_{ij}=-\epsilon_{ji}$ and $\epsilon_{12}=1$) and find that there are two different values for most of our classes. Once again, the latent representation reveals the symmetry structure of the problem. As a consistency check we find that no such structure is observed on the input data.

\subsection{Discrete Case: Identifying distinct string theory vacua}
\label{sec:cicys}
After these warm-up exercises we now discuss an example where finding the symmetries in a dataset are crucial to answer a question in mathematical physics: {\it How many distinct vacua of string theory can be constructed in a particular class of string models?}

Knowing which distinct ways one can obtain string vacua is a crucial question in our understanding of string theory as a theory of quantum gravity. One sub-question is associated to classifying consistent background geometries for string theory, in particular CY-manifolds~\cite{Candelas:1985en}.

CICYs provide an interesting class of such backgrounds: their classification has been achieved in three and partially in four dimensions \cite{Candelas:1987kf,Gray:2013mja} and models on such spaces are among the most realistic string vacua constructions to date~\cite{Anderson:2011ns,Anderson:2012yf}. The initial enumeration features many representations which are related by a priori unknown symmetries. Although they have been identified in a heroic effort for three and four dimensions, it is unknown what the symmetries are in higher dimensions. The knowledge of these symmetries is necessary in order to tackle the combinatorial complexity of the initial enumeration which renders a classification in higher dimensions currently unfeasible.

CICYs are realized as complete intersections in products of complex projective spaces whose classical description we now review (cf.~\cite{Hubsch:1992nu} for more details).

\subsubsection*{Construction -- classical description}
A CICY can be described by its configuration matrix which, for instance, can look like this
\begin{equation*}
\begin{split}
\left[
    \begin{array}{c|cc}
		 1 & 1 & 1 \\
		 2 & 1 & 2 \\
		 3 & 0 & 4
    \end{array}
    \right]\,.
\end{split}
\end{equation*}
The notation is to be understood as follows:
The first column of the matrix denotes the dimension of the projective space, here our space is the product space $\mathbb{P}^1 \times\mathbb{P}^2 \times\mathbb{P}^3$. The other columns encode the information on the polynomials which define the hypersurface in the ambient product space. The entries in a given column refer to the multi-degrees in the corresponding projective space. The CICY is defined as the zeros of these polynomials. To write the polynomials explicitly for this example, we have to define the coordinates of each space: $\mathbb{P}^1$ is denoted with $x^a,$ were $a=0,1$, the $\mathbb{P}^2$ coordinates by $y^{i}$ with $i=0,1,2$, and for $\mathbb{P}^3$ we have $z^m$ with $m=0,1,2,3$. The polynomials can be written as (before imposing any scaling of the projective spaces):
\begin{equation*}
\begin{split}
    p_1=&\sum_{\substack{a=0,1\\ i =0,1,2 }} c_{ai}x^ay^i= c_{00}x^0y^0+c_{01}x^0y^1+c_{02}x^0y^2+c_{10}x^1y^0+c_{11}x^1y^1+c_{12}x^1y^2~,\\
    p_2=&\sum_{\substack{a=0,1\\ i,j=0,1,2 \\
    m,...,q=0,...,3}}d_{aijmnpq}x^ay^iy^jz^mz^nz^pz^q\,,
\end{split}
\end{equation*}
where $c_{ai}$ and $d_{aijmnpq}$ are complex coefficients. Therefore, the configuration matrix describes a family of CICYs parametrised by the space of the coefficients. Many basic properties do not depend on the explicit form of the polynomials, but only on the configuration matrix (so for example the Euler characteristic depends on the configuration matrix rather than on the explicit polynomials). This feature is the strength of this notation, and one of the motivations to introduce it.
For the hypersurface to be a CY-manifold, the rows have to satisfy the following relation between the degree of the projective factor and its appearance in all polynomials:
\begin{equation}
n+1=\sum_\alpha q_n^\alpha~.
\end{equation}

Restricting to manifolds of fixed complex dimension $d$ leads to the constraint on the number of projective components
\begin{equation}
\sum_r n_r=k+d~,
\end{equation}
where $k$ denotes the number of equations. In combination with the observation that a $\mathbb{P}^1$ factor with a quadratic constraint is redundant, it can then be shown that there is only a finite number of such configuration matrices~\cite{Green:1986ck}. In~\cite{Candelas:1987kf} $7890$ of such matrices were singled out for the case of threefolds, utilising some additional identities which are discussed below. This dataset can be found online~\cite{cicylist}. In~\cite{Anderson:2008uw} it was pointed out that $435$ of these matrices are redundant and describe the same CICY. For fourfolds $921,497$ configuration matrices were obtained in~\cite{Gray:2013mja} and in higher dimensions the corresponding sets of configuration matrices are unknown. In the following we focus on the case of three-folds.

\subsubsection*{Identities -- discrete symmetries}

The simplest identities which leave the underlying CICYs unchanged are permutations of rows and columns in the configuration matrices.

Beyond this, there are several further identities how configuration matrices are linked to each other which can be checked explicitly for small configuration matrices and the identities can then be applied in general~\cite{Candelas:1987kf}. To obtain the classification one can choose one of these respective representations. They can be summarised as follows:
\begin{equation}
\label{eq:cicyidentities}
\begin{split}
%first
\left[
    \begin{array}{c|cc}
		 2 & 2 & \bold{a} \\
		 \bold{n} & 0 & \bold{q} \\
    \end{array}
    \right]
    =&
\left[
    \begin{array}{c|c}
		 1 & 2\bold{a}  \\
		 \bold{n} &\bold{q}  \\
    \end{array}
    \right],  \qquad  
%second
\left[
    \begin{array}{c|cc}
		 1 & 1 & \bold{a} \\
		 1 & 1 & \bold{b} \\
		 \bold{n} & 0 & \bold{q} \\
    \end{array}
    \right]
    =
\left[
    \begin{array}{c|c}
		 1 & \bold{a}+ \bold{b}   \\
		 \bold{n}  & \bold{q}  \\
    \end{array}
    \right], \qquad    
%    \\[12px]
%third
\left[
    \begin{array}{c|cc}
		 3 & 2 & \bold{c} \\
		 \bold{n} & 0 & \bold{q} \\
    \end{array}
    \right]
    =
\left[
    \begin{array}{c|c}
		 1 & \bold{c}  \\
   	 1 & \bold{c}  \\
		 \bold{n}  & \bold{q}  \\
    \end{array}
    \right],
\\[12px]   
%fourth		 
\left[
    \begin{array}{c|cc}
		 1 & 2 & 0 \\
		 2 & 1 & \bold{c} \\
		 \bold{n} & 0 & \bold{q} \\
    \end{array}
    \right]
    =&
\left[
    \begin{array}{c|c}
		 1 & \bold{c}  \\
   	 1 & \bold{c}  \\
		 \bold{n}  & \bold{q}  \\
    \end{array}
    \right],\qquad %\\[12px]
   %fifth
\left[
    \begin{array}{c|ccc}
		 2 & 2 & 1 & 0 \\
		 2 & 1 & 1 & \bold{a} \\
		 \bold{n} & 0 & 0 & \bold{q} \\
    \end{array}
    \right]
    =
\left[
    \begin{array}{c|cc}
		 1 & 2 & 0 \\
   	 2 & 2 & \bold{a}  \\
		 \bold{n} & 0 & \bold{q}  \\
    \end{array}
    \right]    .     	 
\end{split}
\end{equation}
Here ${\bf n}$ denotes a vector containing the dimensions of `arbitrary' projective spaces. ${\bf a,~b}$ denote vectors containing zeros everywhere but in one entry which equals one. ${\bf c}$ denotes vector with cross sum two. ${\bf q}$ are appropriate matrices to render the configuration matrix consistent.

\subsubsection*{CICYs as graphs -- new data representation}
The representation in terms of configuration matrices is not permutation invariant, although we are interested in properties which are insensitive to the choice of permutation. This can be achieved when considering a graph representation of the configuration matrix. Such mappings to graphs have shown improved performance such as in classifying properties of molecules~\cite{Kearnes2016}.

\begin{figure}[t]
    \begin{center}
        \begin{tabular}[c]{p{0.3\textwidth}p{0.3\textwidth}p{0.3\textwidth}}
        {\bf Configuration matrix} & {\bf Graph representation} & {\bf Next neighbours}\\
            $\displaystyle
               \left[
	\begin{array}{c|cc}
 		1 & 1 & 1 \\
 		2 & 1 & 2 \\
 		3 & 0 & 4 
	\end{array}
	\right]
            $
            &
            $\vcenter{\hbox{\includegraphics[scale=0.2]{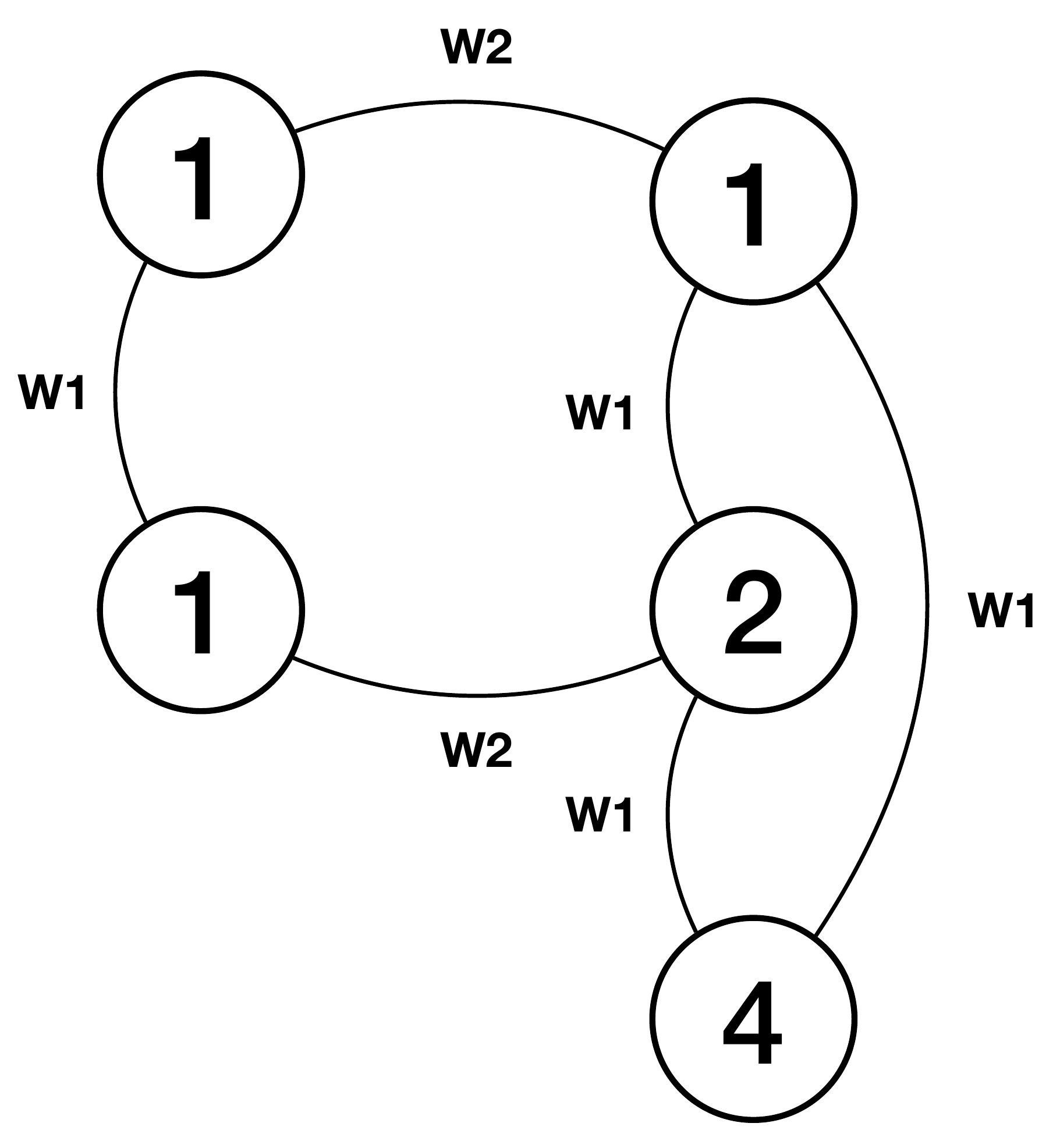}}}$
            &
            \renewcommand{\arraystretch}{1.2}
            $\displaystyle\begin{array}{c |c |c}
            {\rm Vertex} & {\rm horizontal} & {\rm vertical}\\ \hline
            1 & 1 & 1\\
            1 & 1 & 2,4\\
            1 & 2 & 1\\
            2 & 1 & 1,4\\
            4 & - & 1,2
            \end{array}$
        \end{tabular}
\end{center}\vspace{-0.5cm}
\caption{Different representation of one CICY. {\bf Left:} The classic configuration matrix. {\bf Middle:} A graph visualisation with two distinct weights. {\bf Right:} Nearest neighbours of the graph.}\label{fig:graphexample}
    \end{figure}

For this novel representation of CICYs we mapped the right part of the configuration matrix (which is sufficient to reconstruct the whole matrix) to a graph. An example of such a graph is shown in Figure~\ref{fig:graphexample}. We assign different weights to connections in rows and columns respectively. This representation has the advantage that our notation of CICYs is invariant under the permutation of rows and columns.

As the next step, we have to prepare the data in such a way that we can feed the graphs in our network. Therefore, we have to translate the properties of the graphs into a numerical description.
We use the next neighbours of each point which are shown for our example in Figure~\ref{fig:graphexample} on the right side. We calculated these features for all CICYs and hence obtained a dictionary for all types of points in this dataset, finding $285$ types. This naturally gives a $285$-dimensional feature vector with integer entries. As these feature vectors do not uniquely identify a CICY we also use the eigenvalues of the adjacency matrix of the graph as input.
In summary, we took the feature vector which has a clear length consisting of integers and the eigenvalues of the adjacency matrix, padded with additional zeros as input for our network. This leads to a $315$-dimensional input vector.
Note that the identities correspond to local operations on our graphs.

\subsubsection*{Training of the network}
Our target output data are the topological invariants $h^{1,1}$ and $h^{1,2}$ which were obtained in~\cite{Green:1987cr}. For this supervised learning task, we now proceed as in the continuous case, in particular as in the $SU(2)$ case with two output classification layers, one for $h^{1,1}$ and one for $h^{1,2}.$ 

We started from the classified input-output pairs, and constructed $500$ random representatives of each class using identities (if applicable) and permutations. As next step, we constructed the $315$-dimensional input vector as previously described. We note that in this representation each class has a different number of representatives, depending on the number of identities which can be applied. For example the so called quintic hypersurface
\begin{equation}
\begin{split}
\left[
   	 \begin{array}{c|c}
4 & 5 \\
\end{array}
\right]
\end{split}
\end{equation}
just has one representative because no identities can be applied here. However for other CICYs we obtain between $100$ and $300$ representatives. We end up with around $600,000$ different input vectors. The clear advantage of this input is that we can be sure that two different data-points always describe two distinct matrices which are not related via permutations. To balance the discrepancy of different number of representatives we keep several copies of CICYs with low number of representatives in our training dataset. For evaluation of the classification we only use unique input vectors.

The network we use is a simple multilayer-perceptron with ReLu-activation functions and two softmax-classifications as the final layer, details can be found in Table~\ref{tab:nn}. Again we stop training when the network achieves above 95 percent accuracy in both classifications. For the analysis of the results, we only use the correctly classified data-points.

\begin{table}
\begin{center}
 \begin{tabular}{l c l l l}
 Type & Dimension & Activation & Initializer & Regularization\\ \hline
  Input & 315\\
  Dense & 315 & ReLU & \texttt{glorot\_uniform} & \\
  Dense & 315 & ReLU & \texttt{glorot\_uniform} & l2($10^{-5}$)\\
  BatchNormalization\\
  Dense & 100 & ReLU & \texttt{glorot\_uniform} & \\
  Dense & 100 &  & \texttt{glorot\_uniform} & l2($10^{-3}$)\\
  Output 1: Dense & 102 & softmax\\
  Output 2: Dense & 20 & softmax
 \end{tabular}
\end{center}
\vspace{-0.3cm}
\caption{Neural network architecture for Hodge number classification. The embedding layer is the layer before the output layers. We use categorical crossentropy as the loss on both output layers.}
\label{tab:nn}
\end{table}

\subsubsection*{Analysis of the results}

As we face a situation with too many classes we utilise a different method to analyse the nearest neighbours in the embedding layer. For a given input configuration, we look at distances of its nearest neighbours in the embedding layer. We identify a sufficient threshold and compare the class labels of the points closer than the threshold.\footnote{There is no obstruction to apply this procedure also in the previous situations.}

\begin{figure}[t]
\begin{center}
\includegraphics[width=0.49\textwidth]{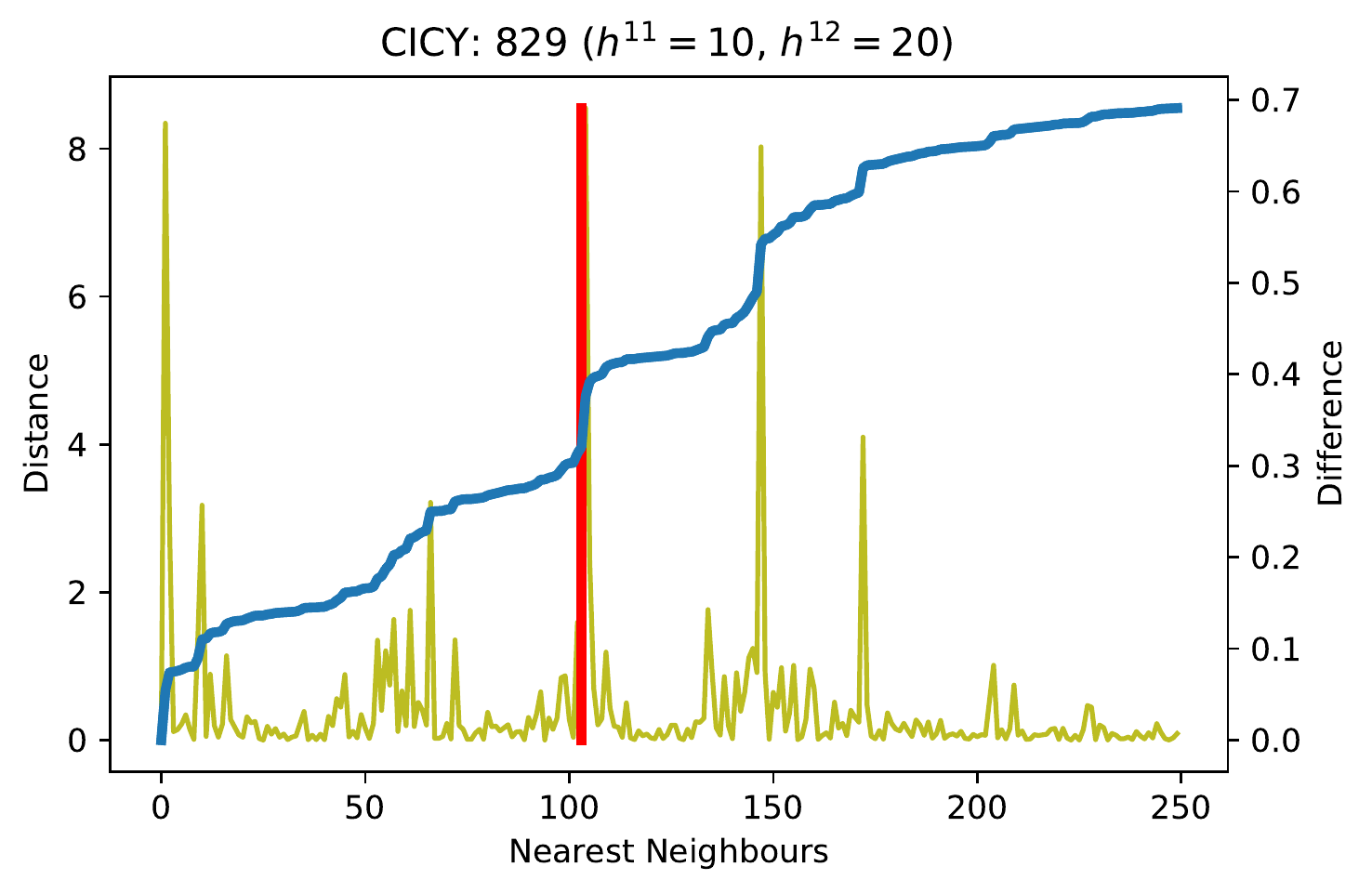}
\includegraphics[width=0.49\textwidth]{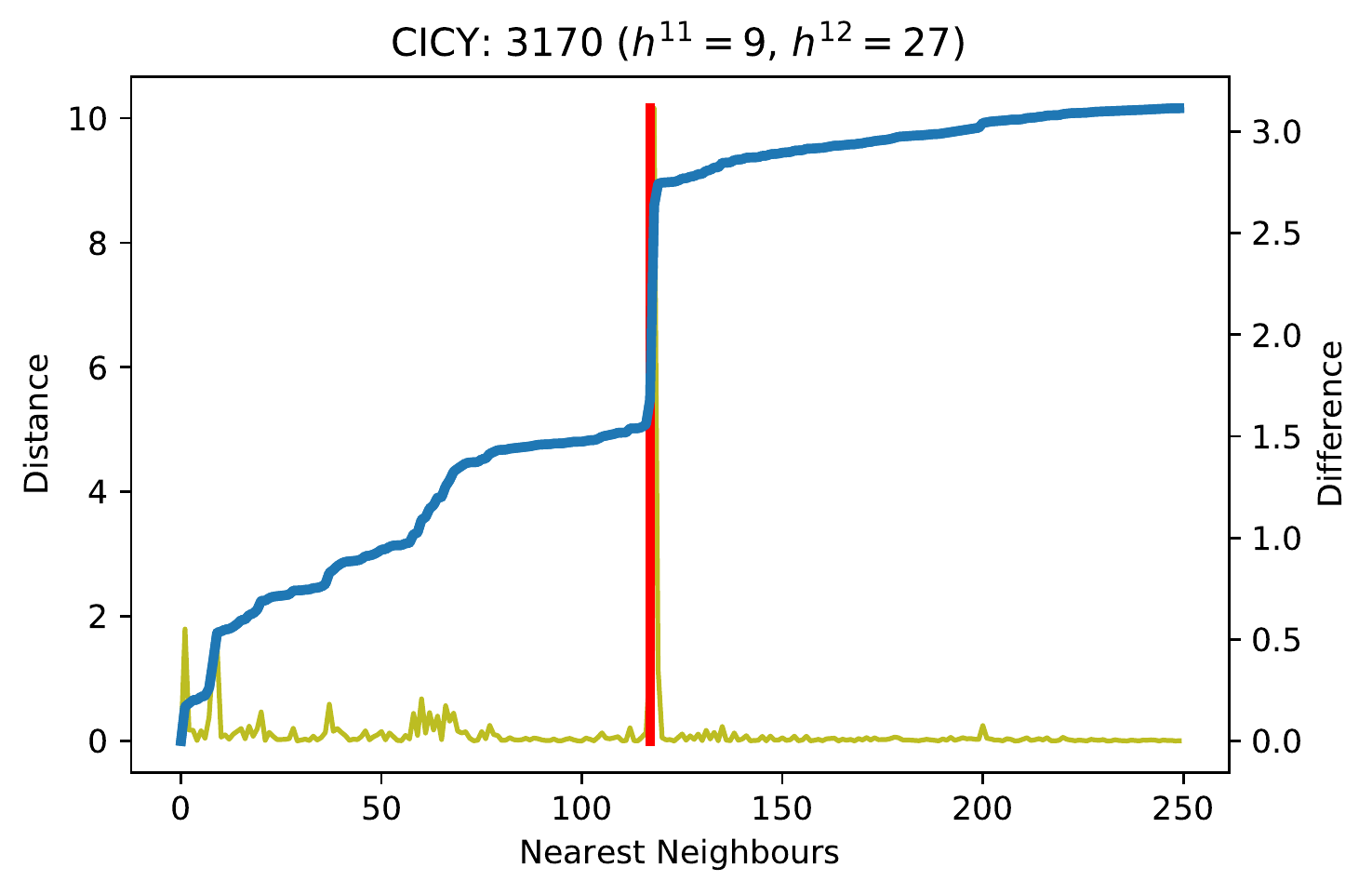}
\adjustbox{margin=0em,vspace=-1.0cm,width=0.85\textwidth,set height=2.5cm,set depth=2.cm,center}{{\footnotesize $\left[\begin{array}{c|c c c c c c c c c c}
1 & 1 & 1 & 0 & 0 & 0 & 0 & 0 & 0 & 0 & 0\\
1 & 0 & 0 & 1 & 1 & 0 & 0 & 0 & 0 & 0 & 0\\
1 & 0 & 0 & 0 & 0 & 1 & 1 & 0 & 0 & 0 & 0\\
1 & 0 & 0 & 0 & 0 & 1 & 0 & 1 & 0 & 0 & 0\\
2 & 0 & 0 & 0 & 0 & 0 & 0 & 0 & 1 & 1 & 1\\
2 & 0 & 0 & 0 & 0 & 1 & 0 & 0 & 1 & 1 & 0\\
2 & 1 & 0 & 1 & 0 & 0 & 0 & 0 & 0 & 0 & 1\\
3 & 0 & 1 & 0 & 1 & 0 & 1 & 1 & 0 & 0 & 0
 \end{array}\right]$\hspace{2.0cm} $\left[\begin{array}{c|c c c c c c c c c c}
1 & 1 & 1 & 0 & 0 & 0 & 0 & 0 & 0 & 0 & 0\\
1 & 0 & 0 & 1 & 0 & 0 & 0 & 0 & 1 & 0 & 0\\
1 & 0 & 0 & 0 & 1 & 0 & 0 & 1 & 0 & 0 & 0\\
1 & 0 & 0 & 0 & 0 & 1 & 1 & 0 & 0 & 0 & 0\\
1 & 0 & 0 & 0 & 0 & 0 & 0 & 0 & 0 & 2 & 0\\
2 & 0 & 0 & 0 & 0 & 0 & 1 & 0 & 1 & 1 & 0\\
2 & 1 & 0 & 0 & 0 & 0 & 0 & 1 & 0 & 1 & 0\\
2 & 0 & 1 & 0 & 0 & 1 & 0 & 0 & 0 & 0 & 1\\
2 & 0 & 0 & 1 & 1 & 0 & 0 & 0 & 0 & 0 & 1 
 \end{array}\right]$}}

\end{center}\vspace{-1.0cm}
\caption{We show the Euclidean distance of the $250$ nearest neighbours in the embedding layer to two fixed CICYs (blue). In yellow we show the difference between these distances for points $i$ and $i+1.$ In red we highlight the largest difference. Below is the respective CICY configuration matrix from the original list.\label{fig:distanceembedding}}
\end{figure}

\begin{figure}[t]
\includegraphics[width=0.49\textwidth]{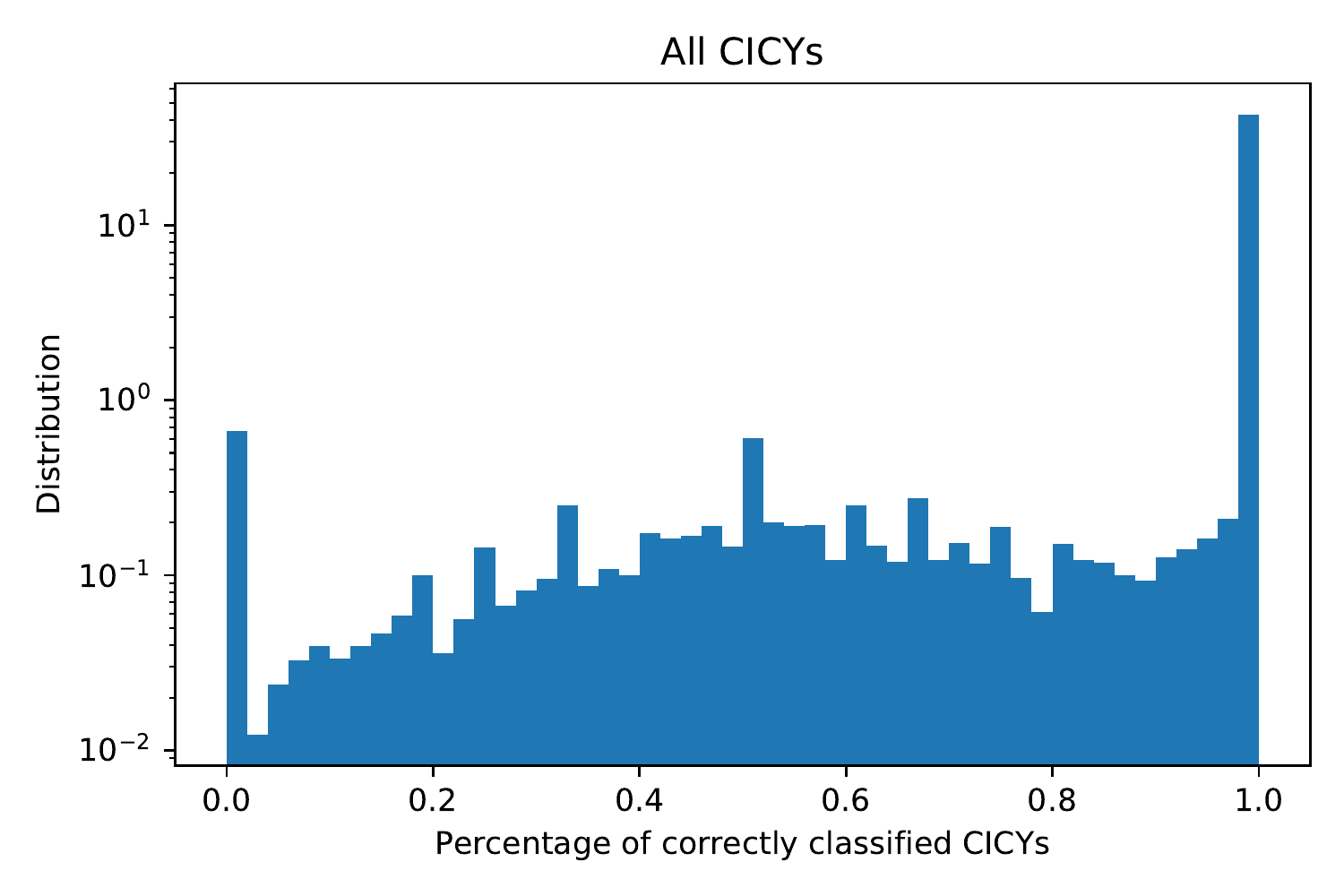}
\includegraphics[width=0.49\textwidth]{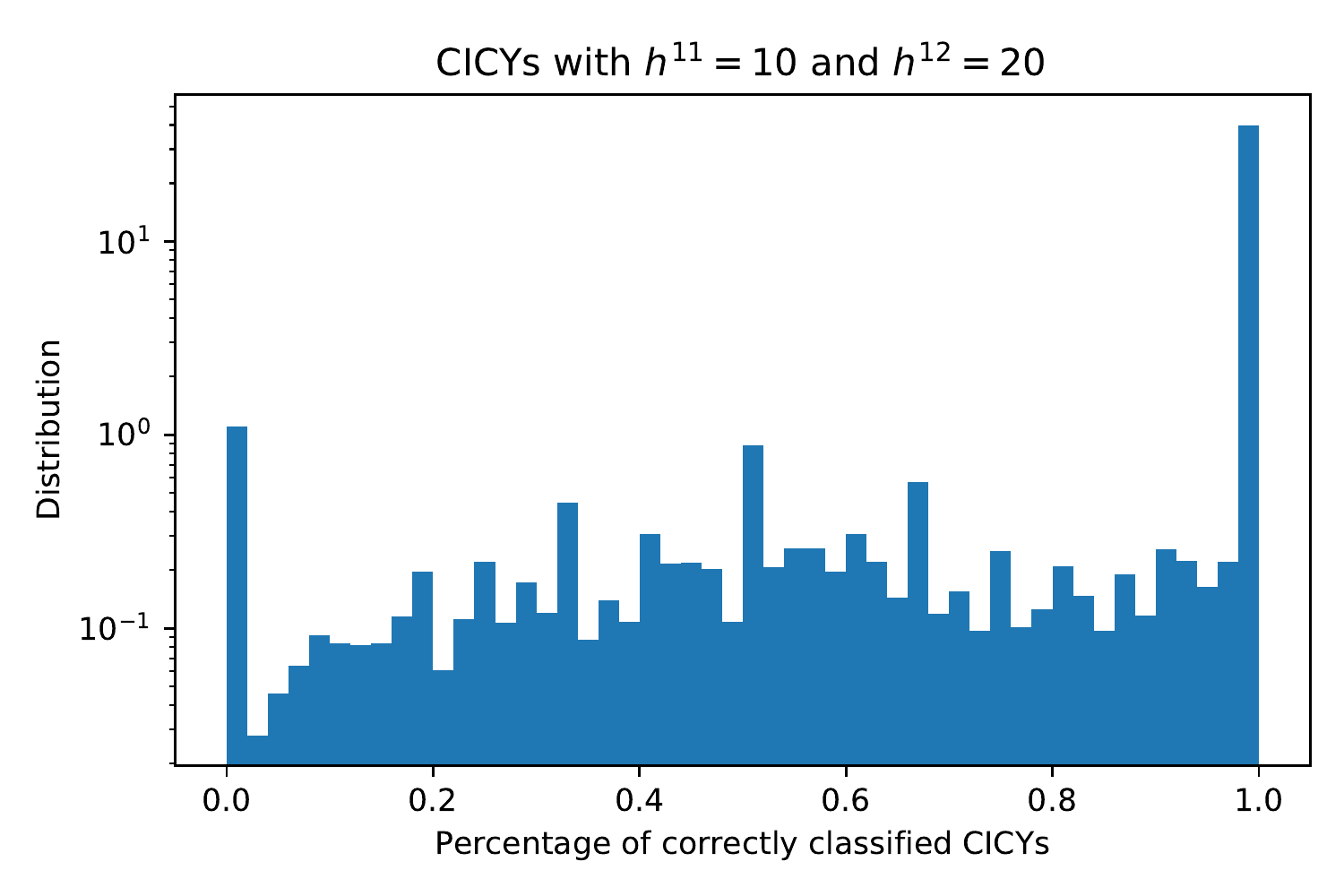}
\vspace{-0.1cm}
\caption{Performance of our method on CICY dataset. {\bf Left:} The distribution of performance for all 686,464 data points. {\bf Right:} The distribution of performance on the subset of CICYs with Hodge-numbers $h^{1,1}=10$ and $h^{1,2}=20$. The analysis of finding nearest neighbours is still performed with all data points.}
\label{fig:cicyhist}
\end{figure}

As a first step, we pick one data point in the embedding space and find the 250 nearest neighbours with respect to their Euclidean distance. A plot of these lines are the blue curves in Figure~\ref{fig:distanceembedding}.
Two generic features are several plateaus in the distance curve and several big jumps between two points which are shown in yellow in Figure~\ref{fig:distanceembedding}. We are interested in the biggest jump, and we use this as our threshold to distinguish manifolds. The red line in Figure~\ref{fig:distanceembedding} is the location of the threshold. The prediction is that points closer than the point at the threshold all belong to one class. We require that we are looking at least at one neighbour. This prediction is quite successful given the fact that the network is just trained with the Hodge numbers, and has no training on the CICY labels. Figure~\ref{fig:cicyhist} summarises the performance of our method with respect to the CICY labels and we find that for the vast majority of data points the neighbours are correctly classified (for $86.6\%$ of CICY labels we find an accuracy above $95\%$). Outliers arise for CICYs with one or two existing representatives which is expected from this method. Focusing on the Hodge pair with $h^{1,1}=10$ and $h^{1,2}=20,$ there are $292$ distinct CICYs. Again (cf.~Figure~\ref{fig:cicyhist} right panel), we find that the majority of the CICYs are correctly classified with our method -- noting only a small drop to $80.6\%$ compared to the performance on the entire dataset. Such a drop is expected because the entire dataset contains many cases where we have just one class of CICYs for a specific combination of Hodge numbers.

The surprising part is that as far as we know there is no straightforward way to see whether two manifolds are inequivalent due to the basis dependence of the intersection numbers. Therefore, more analysis is in order to understand why networks are able to distinguish distinct matrices, and find a sufficient basis to distinguish between CICYs. We plan to return to this question whether the neural network has learned Wall's theorem~\cite{Hubsch:1992nu}.\footnote{We thank Per Berglund and Andre Lukas for stressing this observation to us.}

\section{Finding Generators}
\label{sec:generators}
Having identified the presence of symmetries, the next step, which we discuss now, is to identify the symmetry generators. Our starting point is a pointcloud on the input space which has been identified in the previous step to be related via a symmetry due to the closeness in the embedding layer. To establish a numerical method to perform this analysis we start with a noisy pointcloud. First, we describe our algorithm and apply it then in examples for several symmetry groups in various dimensions. Finally we exemplify how this algorithm can be utilised on images.

\subsection{Algorithm}
The idea behind the algorithm is to extract the information about the symmetry group when considering a pointcloud $P$ which has been found to be related by some symmetry group. Infinitesimally, points are connected as follows:
\begin{equation}
 p'=p+\epsilon_a T^a p~,
 \label{eq:characteristic}
\end{equation}
where $\epsilon_a$ are some small numbers selecting by how much the point is transformed with the respective generator $T^a.$ The symmetry group is characterised by the generators $T^a$ which we want to obtain from the pointcloud. In particular the structure of the nearest neighbours carries the information about the generators. To extract them efficiently, one needs to find an appropriate regression setup where all components of the generators $T^a$ are constrained. For instance, considering just a single point in $n$-dimensions gives via equation~\eqref{eq:characteristic} $n$ conditions on the components. However, by appropriately utilising multiple points the generators can be completely identified. We find the generators as follows:
\begin{enumerate}
 \item If our dataset features several redundant dimensions or the inputs are not centered around the origin to pre-process the dataset by performing appropriate dimensional reduction and centering around the origin (e.g.~via PCA).
 
 \item We generate an orthonormal basis $(b_1,\ldots,b_n)$ as follows. We pick a point $p_1\in P$ at random. The first basis vector is given by its associated normalised vector $b_1=p_1/||p_1||.$ We then pick a further  vector at random in the pointcloud $P,$ and the second basis vector is given by the normalised version of $p_2-(p_2\cdot b_1)b_1.$ We then complete the remaining orthonormal basis elements automatically.

  \item The next step is to filter out points which are close enough to the hyperplane $H$ spanned by $b_1$ and $b_2$. This is the hyperplane in which the generator acts. As condition we use
  \begin{equation}
  |p\cdot b_i|< \delta~\text{ for } 2<i\leq n~. 
 \end{equation}
 The more data points we have the smaller we can choose $\delta$. Points in this `thick' hyperplane  feature neighbours in the direction of interest and points in the orthogonal direction. The contribution of these latter points to our regression problem is removed later with condition~\eqref{eq:c2}. Note that a too large $\delta$ will include all points -- in particular also the poles on the sphere -- which leads to a drop in performance.

  \item Within this points we now identify all pairs of points  $p,p' \in H$ which are close to each other:
 \begin{equation}
   ||p- p' ||< \epsilon~\text{ for } \forall\, p,p'\in H~. 
\end{equation}
This choice allows us to keep multiple point pairs and not just the nearest neighbour.

\item Each of these neighbouring point pairs $(p,p')$ provides constraints relevant for determining one combination of the generators in Equation~\eqref{eq:characteristic}. At linear order this is given as
\begin{equation}
 p'-p=\frac{\sigma_H(p,p')}{\lVert p\rVert}~\lVert p'-p\rVert~T p~,
\end{equation}
where $T$ denotes the generator we determine. The normalisation factor $1/\lVert p\rVert$ ensures the correct numerical prefactors. $\sigma_H(p,p')$ denotes the sign which contains the appropriate directional information of the points $(p,p')$ for this hyperplane and is calculated by
\begin{equation}
 \sigma_H(p,p')=~{\rm sign}\left(
 (p\cdot b_1)(p'\cdot b_2)-(p\cdot b_2)(p'\cdot b_1)
 \right).
 \label{eq:c1}
\end{equation}
The necessity of $\sigma$ can be understood by considering the example of identifying the generator of $SO(2)$ and considering point pairs in different quadrants. Each of these point pairs constrains up to $n$ components of the $n\times n$-components of $T.$ Additional components are constrained by demanding that
\begin{equation}
 T~b_i=0~\text{ for }i>2~.
 \label{eq:c2}
\end{equation}
\item Using the above constraints in Equations~\eqref{eq:c1} and~\eqref{eq:c2} we now can constrain all components of the generator using linear regression. In practice we weigh the constraints arising from~\eqref{eq:c2} stronger than constraints from~\eqref{eq:c1}, ensuring that~\eqref{eq:c2} is definitely satisfied. This also removes the false directional information arising from point pairs arising due to the thickness of our hyperplane.
\item By applying steps 2-5 multiple times we obtain generators for `all' directional combinations. On the resulting generator candidates we perform principal component analysis. By analysing the standard deviation in these components we identify the relevant number of generators for the underlying pointcloud. The associated principal components to these generators reveal the algebra structure of these generators. Hence we determine the underlying symmetry group.

\item To distinguish unitary from orthogonal groups such as in the example below where we distinguish between $SU(2)$ and $SO(4)$ additional care is needed in setting up the regression problem. The necessity arises as follows: Consider the orbit of a point on a unit sphere $S^3.$ The entire orbit which is generated by both symmetries is given by $S^3$ and hence one cannot distinguish with just one pointcloud. However realistic situations such as the example with the $SU(2)$ superpotential (cf.~Section~\ref{sec:symmetries}) feature multiple orbits, one for each field. We can utilise this situation as we are equipped with two point pairs which are connected with the same transformation (neglecting for the moment that they can be in different representations). Here one can distinguish the transformations from $SU(2)$ and $SO(4)$ as the action on the first point pair fixes the $SU(2)$ generator completely, whereas for $SO(4)$ not all generators are fixed by the first transformation. Utilising both point pairs in our regression doubles the constraints arising from~\eqref{eq:c1} and allows us to distinguish for instance $SU(2)$ and $SO(4).$
\end{enumerate}
Below, we discuss some numerical examples of these generators.
\begin{figure}
\begin{center}
 \includegraphics[width=0.3\textwidth]{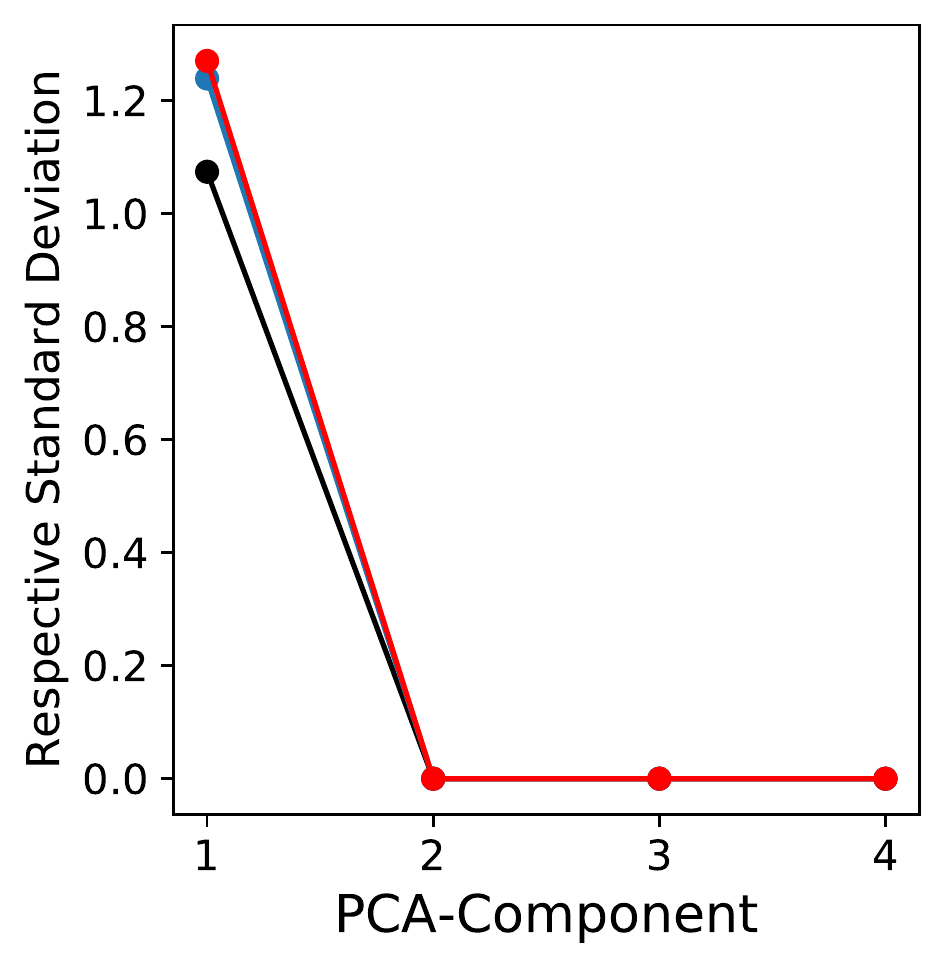}
\raisebox{0.2\height}{\includegraphics[width=0.22\textwidth]{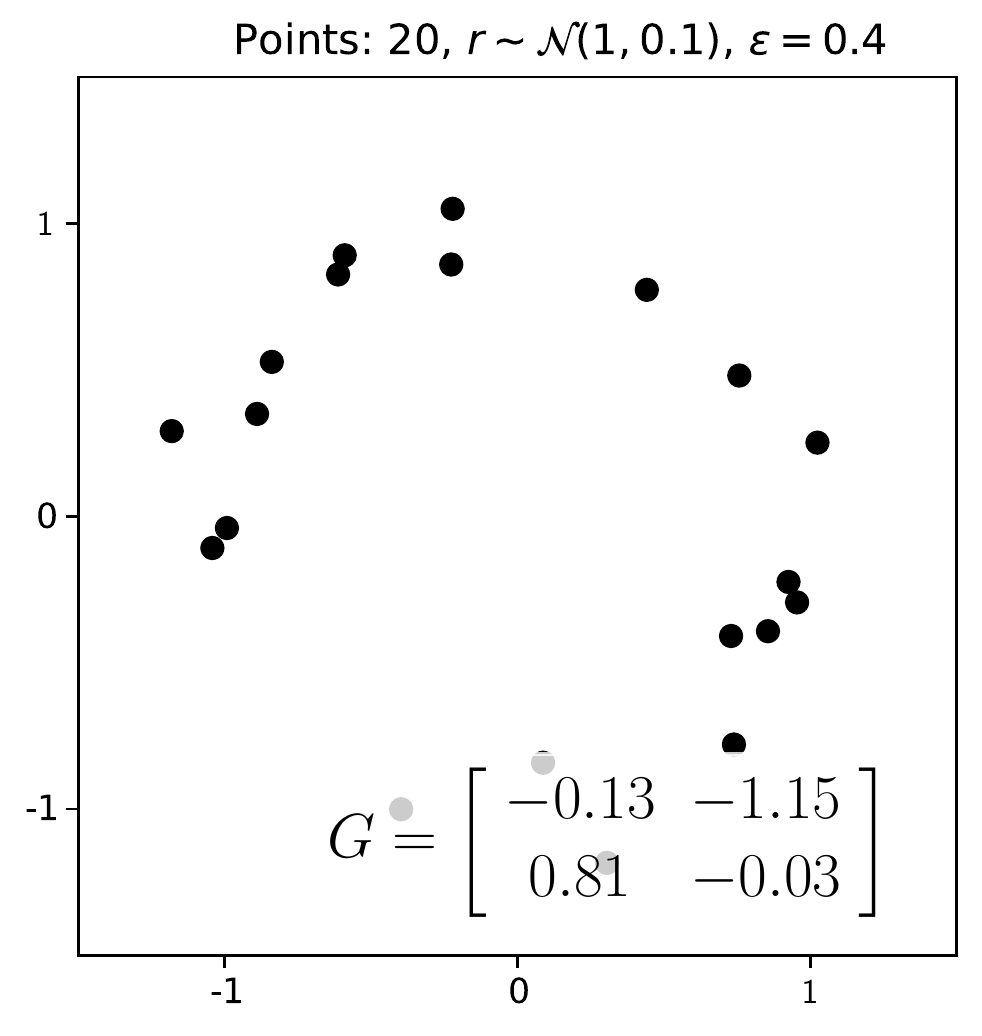}}
\raisebox{0.2\height}{\includegraphics[width=0.22\textwidth]{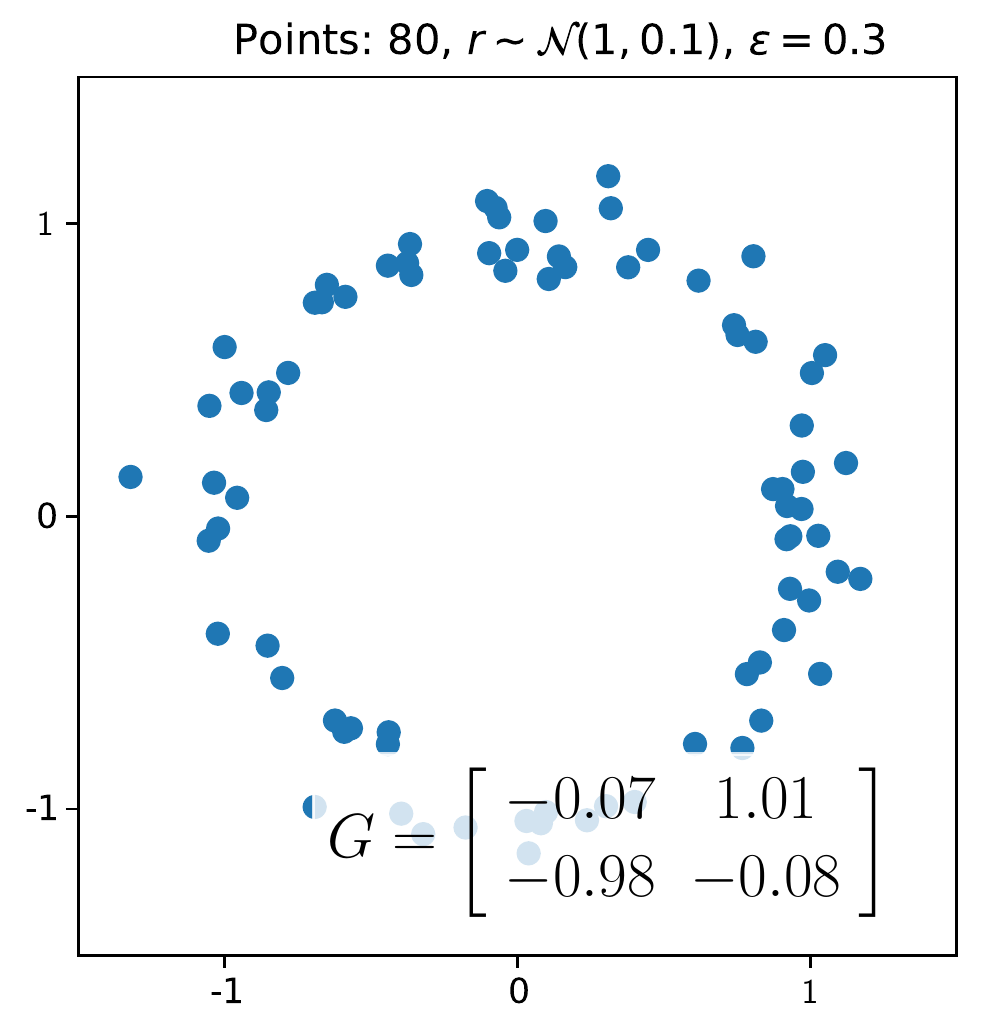}}
\raisebox{0.2\height}{\includegraphics[width=0.22\textwidth]{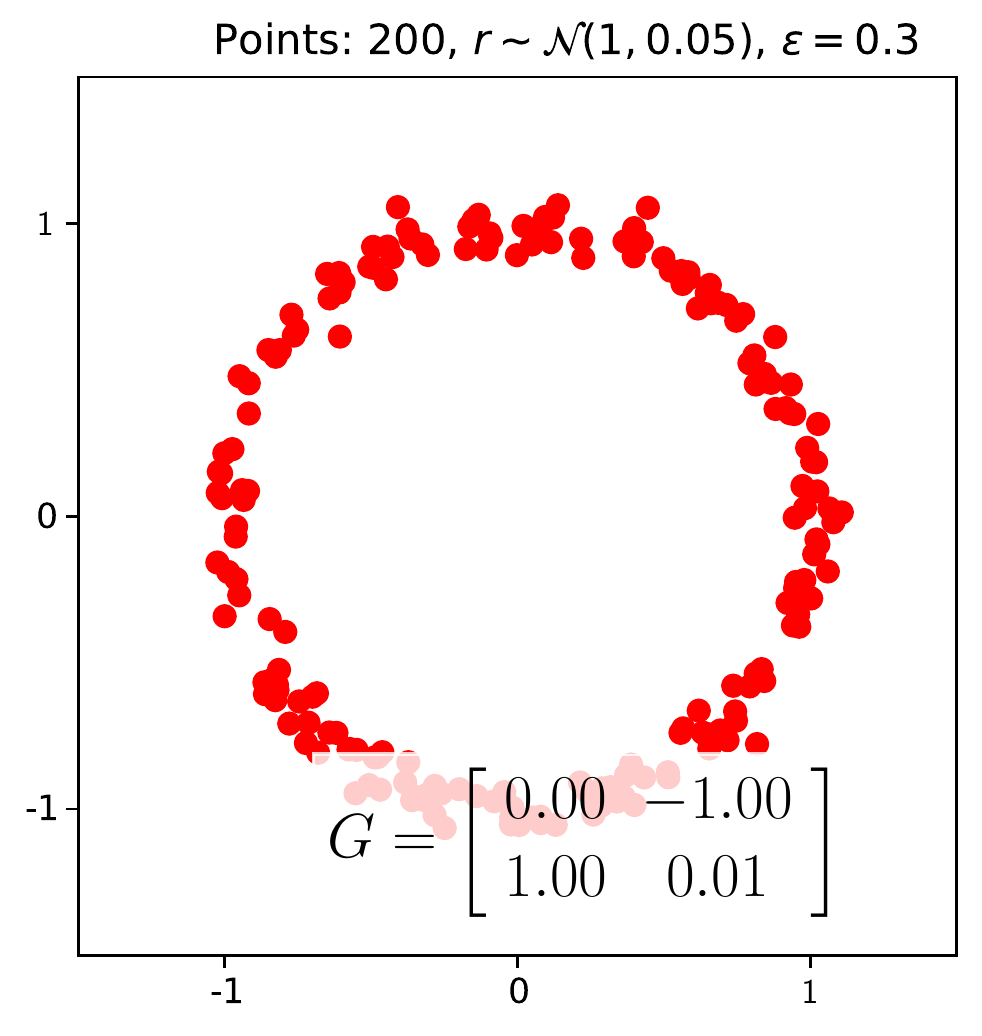}}

\end{center}
\vspace{-0.5cm}
 \caption{Three examples of pointclouds for $SO(2)$ with varying number of points and different noise where the respective parameters are shown in the plot title. The respective generator corresponds to the first PCA component which is singled out by our algorithm.}
 \label{fig:so2}
\end{figure}
\subsection{Examples}
We design our examples in increasing complexity and capture various embeddings of symmetries to check the performance of our algorithm. The first warm-up example is that of a pointcloud generated by $SO(2)$, i.e.~points on a circle. 

To test the stability of our algorithm we perform experiments with varying number of points and we add some Gaussian noise to the radius. Results for several examples are shown in Figure~\ref{fig:so2}. Even for pointclouds with few points and large noise we find very good results for the generators. The large difference in the standard deviation from the first to the remaining components shows that this pointcloud is only connected with one generator. For the analysis shown here we use $\delta=0.5.$

\begin{figure}
\begin{center}
\includegraphics[width=0.47\textwidth]{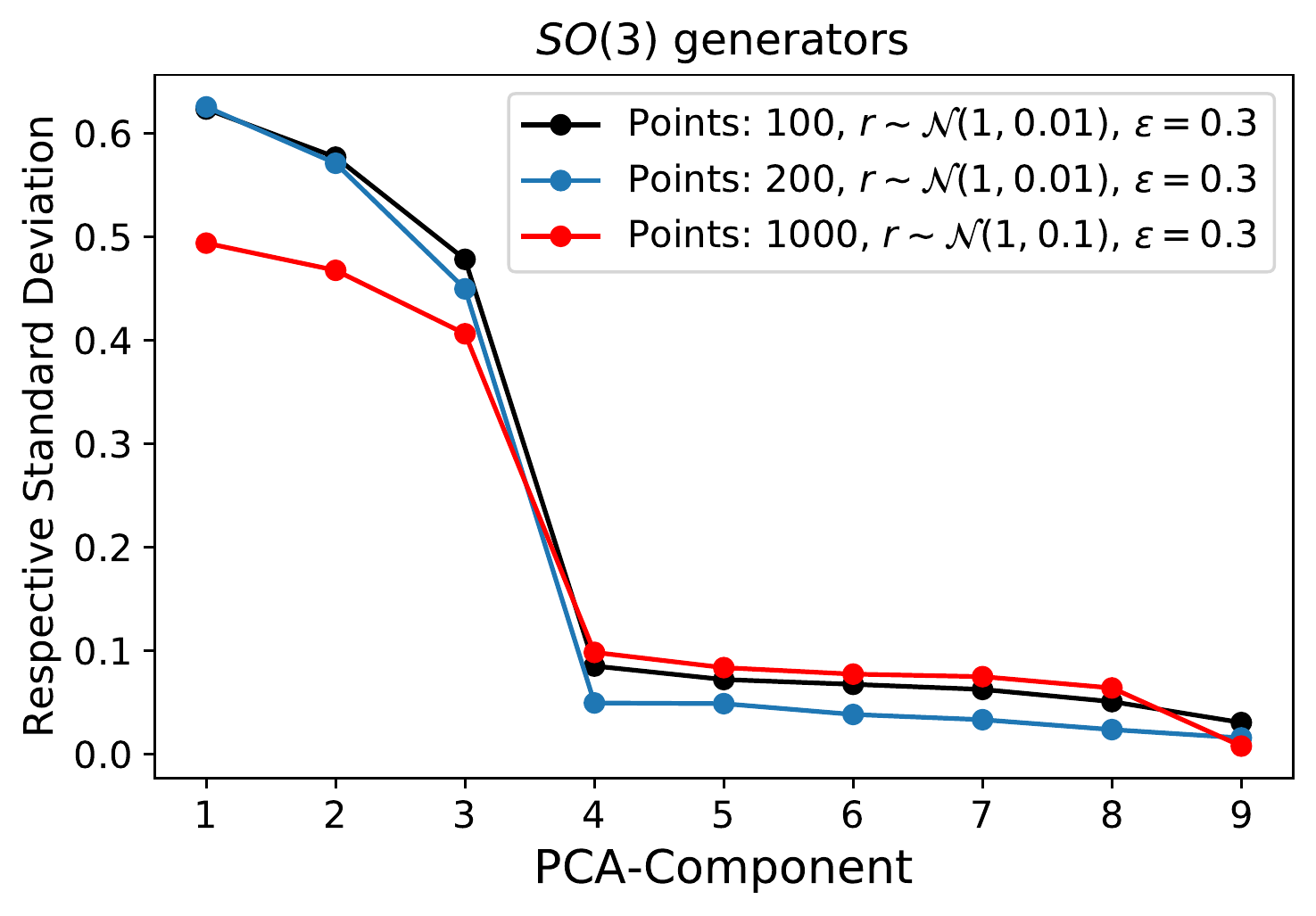}
\includegraphics[width=0.48\textwidth]{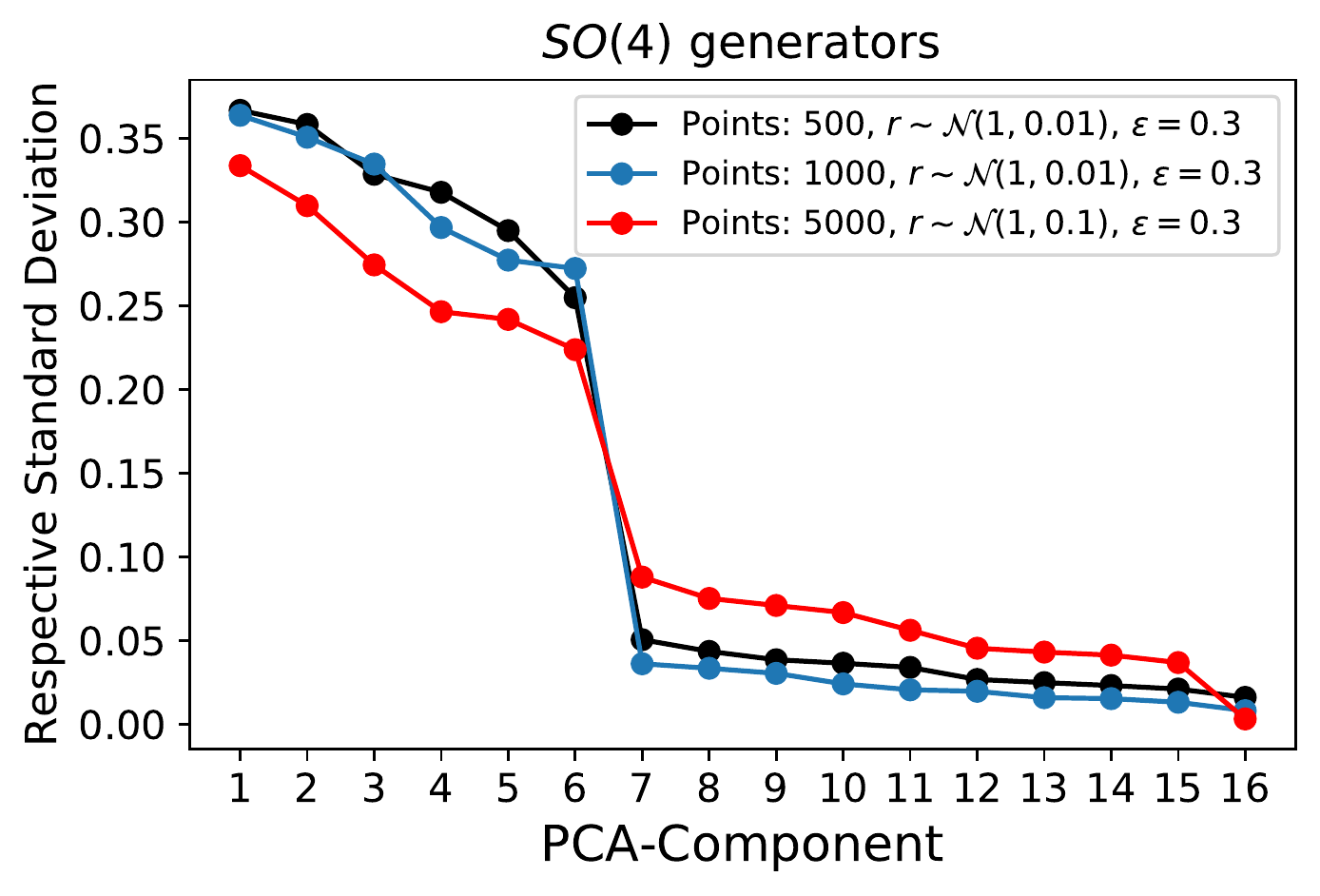}
\end{center}
\vspace{-0.5cm}
\caption{{\bf Left:} The standard deviation of the PCA components for the example of $SO(3).$ {\bf Right:} The results for the standard deviation of the PCA components of the $SO(4)$ example.}\label{fig:so3so4}
\end{figure}

The next examples we discuss are $SO(3)$ and $SO(4).$ Again we train pointclouds with varying total number of points and different levels of noise. For several choices of hyperparameters we show the standard deviations of the PCA-components in Figure~\ref{fig:so3so4}. In both setups we again find consistently a steep decline in the standard deviation after three and six components respectively. For the $SO(3)$ experiment shown as the red curve in Figure~\ref{fig:so3so4} we obtain the following generators:

\begin{equation}
\begin{footnotesize} G_1=\left(
\begin{array}{c c c }
-0.00 &  0.04 & 0.59\\
-0.06 & 0.01 & 0.78\\
-0.59 & -0.82 & -0.01
\end{array}
\right),~
G_2=\left(
\begin{array}{c c c}
-0.01 & -0.98 & -0.13\\
 0.98 & 0.04 & 0.14 \\
 0.16 & -0.18 & 0.01
\end{array}
\right),~G_3=\left(
\begin{array}{c c c}
0.00 & 0.21 & -0.81\\
-0.21 & 0.00 & 0.55\\
 0.78 & -0.61 & -0.03
\end{array}
\right).\end{footnotesize}
\end{equation}
For the $SO(4)$ experiment we obtain the following generators

\begin{eqnarray}
\nonumber
&& \begin{footnotesize}G_1=\left(
\begin{array}{c c c c}
 0.02 &  0.50 & -0.11 & 0.25\\
-0.52 & -0.00 &  0.39 & 0.60 \\
 0.10 & -0.41 & -0.00 & -0.38\\
-0.28 & -0.59 &  0.38 & -0.02
\end{array}
\right),~
G_2=\left(
\begin{array}{c c c c}
 0.00 &   0.08&  0.41& -0.07\\
-0.09 &-0.00 &  -0.31& -0.29\\
-0.48 & 0.24& -0.02& -0.78\\
 0.06 & 0.29&  0.81&  0.02
\end{array}
\right),\end{footnotesize}\\
&&\begin{footnotesize}G_3=\left(
\begin{array}{c c c c}
 0.02&  0.13&  0.42& -0.29\\
-0.09&  0.04&  0.78& -0.32\\
-0.45& -0.76& -0.02&  0.13\\
 0.31&  0.33& -0.13& -0.03
\end{array}
\right),~G_4=\left(
\begin{array}{c c c c}
 0.03&  0.55 & 0.50 &  0.42\\
-0.57& -0.00 &  -0.30 & -0.08\\
-0.45&  0.31 & 0.02 & 0.44\\
-0.40&   0.12& -0.43 & -0.03
\end{array}
\right),\end{footnotesize}\\ \nonumber &&\begin{footnotesize}
G_5=\left(
\begin{array}{c c c c}
 0.01 & 0.63 &-0.48 &-0.50 \\
-0.64 & 0.01 &-0.16 &-0.29\\
 0.48 & 0.14 &-0.00 &   0.03\\
 0.51 & 0.32 & 0.02 &-0.01
\end{array}
\right),~G_6=\left(
\begin{array}{c c c c}
-0.02 & -0.01 & -0.37 & 0.62\\
 0.01 & 0.02 & 0.24 & -0.61\\
 0.40 & -0.24 & -0.03  &-0.13\\
-0.67 & 0.59 & 0.14 & 0.02
\end{array}
\right).\end{footnotesize}
\end{eqnarray}

\begin{figure}
\begin{center}
\includegraphics[width=0.48\textwidth]{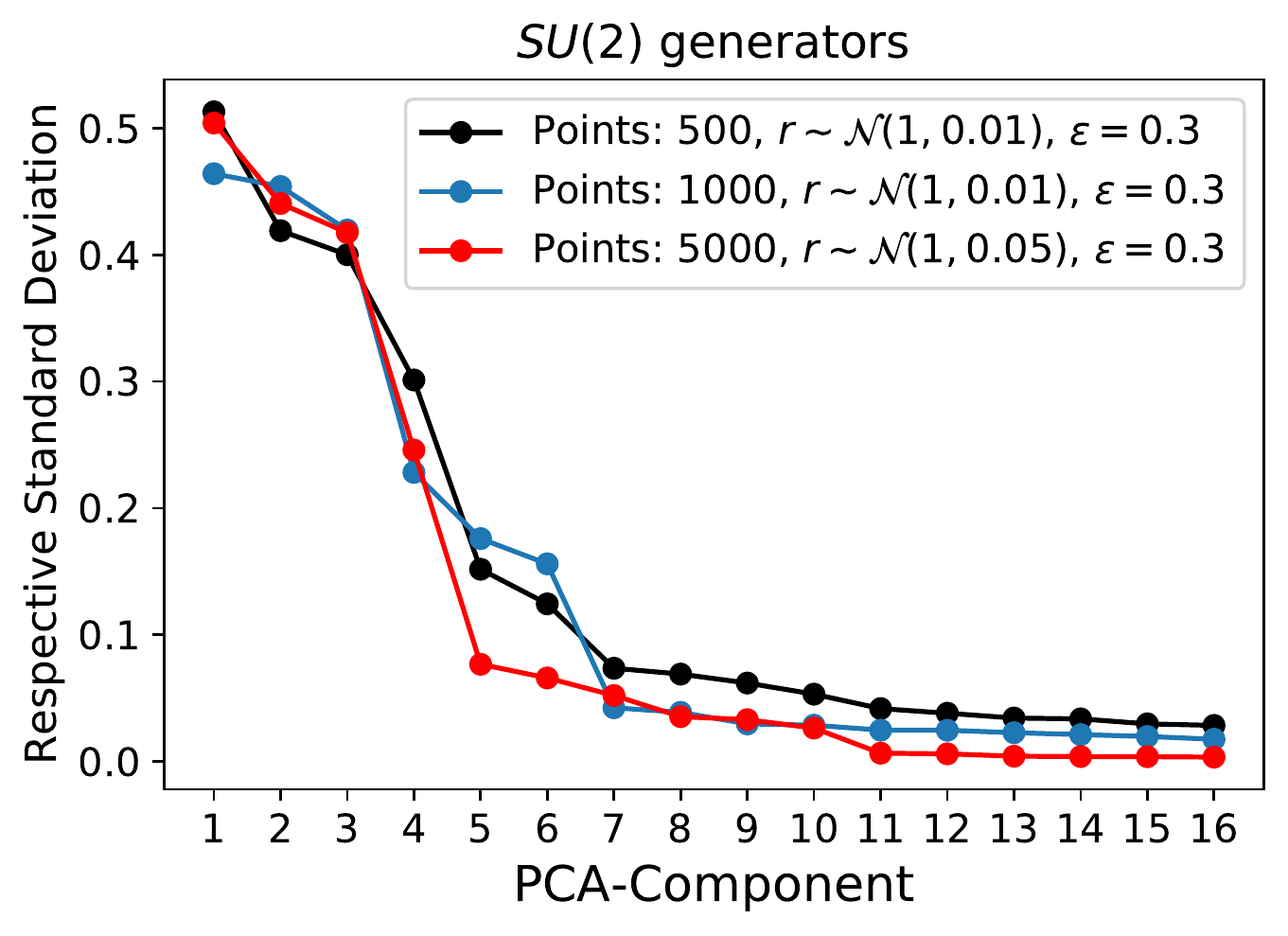}
\includegraphics[width=0.43\textwidth]{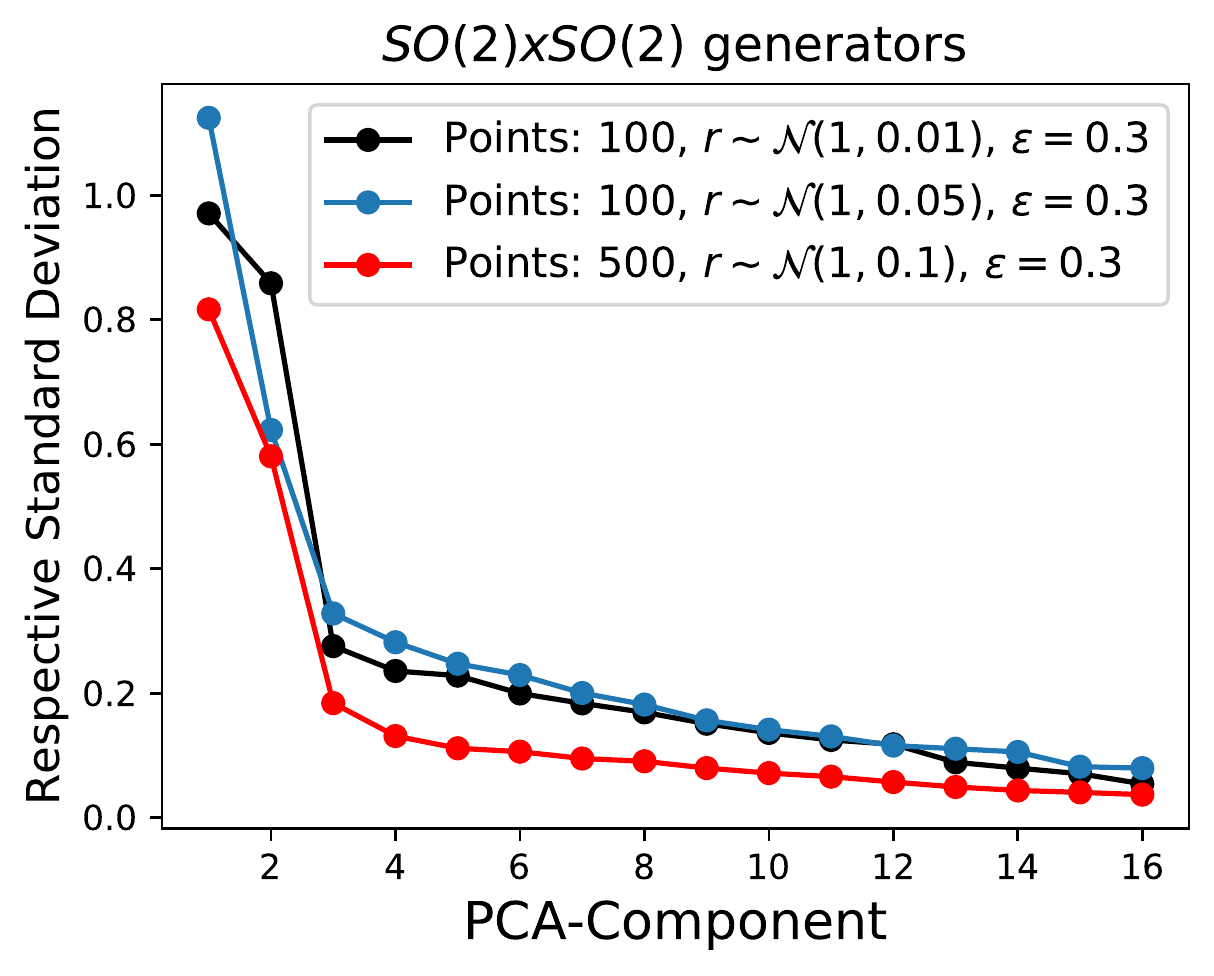}
\end{center}
\vspace{-0.5cm}
\caption{{\bf Left:} The standard deviation of the PCA components for the example of $SU(2).$ {\bf Right:} The results for the standard deviation of the PCA components of the $SO(2)\times SO(2)$ example.}\label{fig:SU2so2so2}
\end{figure}

Next we turn to the discussion of $SU(2)$ and $SO(2)\times SO(2)$ acting on four real dimensions. Our method should reveal the underlying generators three and respectively two generators rather than all six generators of $SO(4).$ Again we test our method on pointclouds with varying number of points and different noise. We provide an overview of our findings in Figure~\ref{fig:SU2so2so2}. For the $SU(2)$ case, the dominant generators found by our algorithm are given:
\begin{eqnarray}
%5000
\nonumber \begin{footnotesize}
G_1\end{footnotesize} &=&\begin{footnotesize}\left(
\begin{array}{c c c c}
-0.01 & 0.52 & 0.47 & -0.11\\
-0.52 & 0.00  &  0.08 & 0.49\\
-0.47 & -0.08 & 0.01 & -0.50 \\
 0.12 &-0.48 & 0.50  & 0.00  
\end{array}
\right),~
G_2=\left(
\begin{array}{c c c c}
-0.00 &  -0.24 & 0.43 & 0.46\\
 0.26 & 0.00  & -0.52 & 0.39\\
-0.43 & 0.51 & -0.00 &   0.35\\
-0.45 & -0.39 & -0.34 & -0.01
\end{array}
\right),\end{footnotesize}\\\begin{footnotesize}G_3\end{footnotesize}&=&\begin{footnotesize}\left(
\begin{array}{c c c c}
 0.00 &  -0.39 & 0.30 & -0.50 \\
 0.37 & 0.01 & 0.51 & 0.32\\
-0.31 & -0.50 &  0.01 & 0.39\\
 0.49 & -0.31 & -0.40 &  0.00  
\end{array}
\right),\end{footnotesize}
\end{eqnarray}
where these results correspond to the run with $5000$ points shown in red in Figure~\ref{fig:SU2so2so2}. Note that to distinguish $SU(2)$ from $SO(4)$ it was necessary to utilise two pointclouds as described in bullet point of our algorithm. For $SO(2)\times SO(2)$ we find, for instance in the case of the run associated to the parameters of the black curve in Figure~\ref{fig:SU2so2so2}
\begin{equation}
\begin{footnotesize}
G_1=\left(
\begin{array}{c c c c}
-0.03 & 0.14 & 0.03 & -0.01\\
-0.31 & 0.01 & 0.03 & 0.01\\
 0.04 & 0.01 &-0.01 & 0.95\\
-0.1 & -0.06 &-0.98 & 0.04
\end{array}
\right),\qquad
G_2=\left(
\begin{array}{c c c c}
 0.00   &-1.13&  0.09 & 0.01\\
 0.78 &-0.04& -0.03 & -0.03\\
 0.03 & 0.04& -0.02 & 0.19\\
 0.08 &-0.02& -0.23 & -0.00  
\end{array}
\right).
\end{footnotesize}
\end{equation}

\subsection{Rotated MNIST}
The final example we discuss is the application of our algorithm on images. To do this we want to re-identify $SO(2)$ from the rotated MNIST dataset $D_{\rm all}.$\footnote{Our rotated MNIST dataset consists of the first $200$ original images in the MNIST dataset and $100$ rotated versions of these images, totalling $20,000$ images. The rotation angles are chosen at random.}
In contrast to our previous examples we now want to identify the generators on a $28\times 28=784$-dimensional space. However, as previously described, we can dimensionally reduce this space, for instance via PCA.

\begin{figure}
 \begin{center}
  \includegraphics[width=0.4\textwidth]{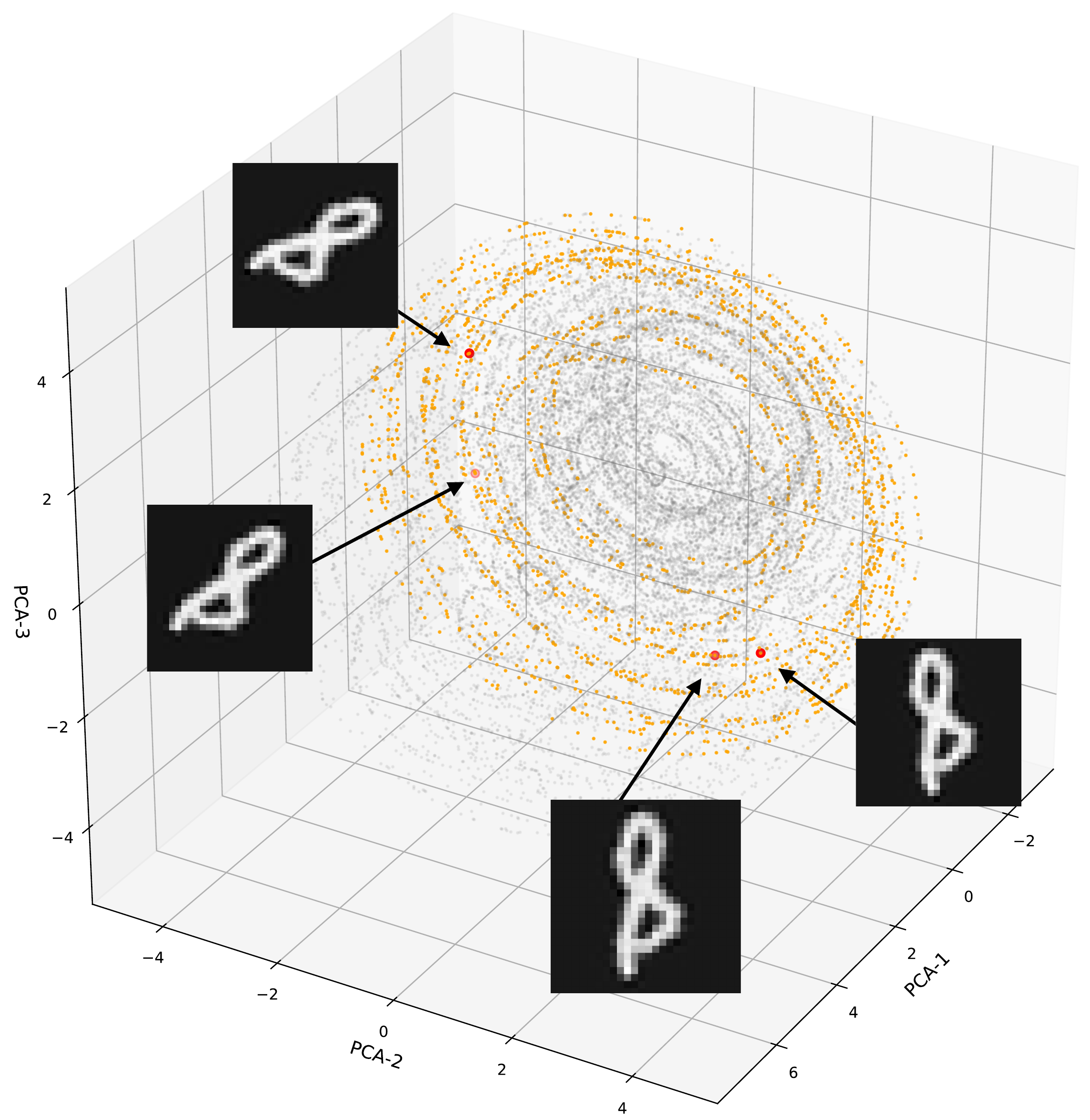}\hspace{0.8cm}
  \raisebox{0.1\height}{\includegraphics[width=0.43\textwidth]{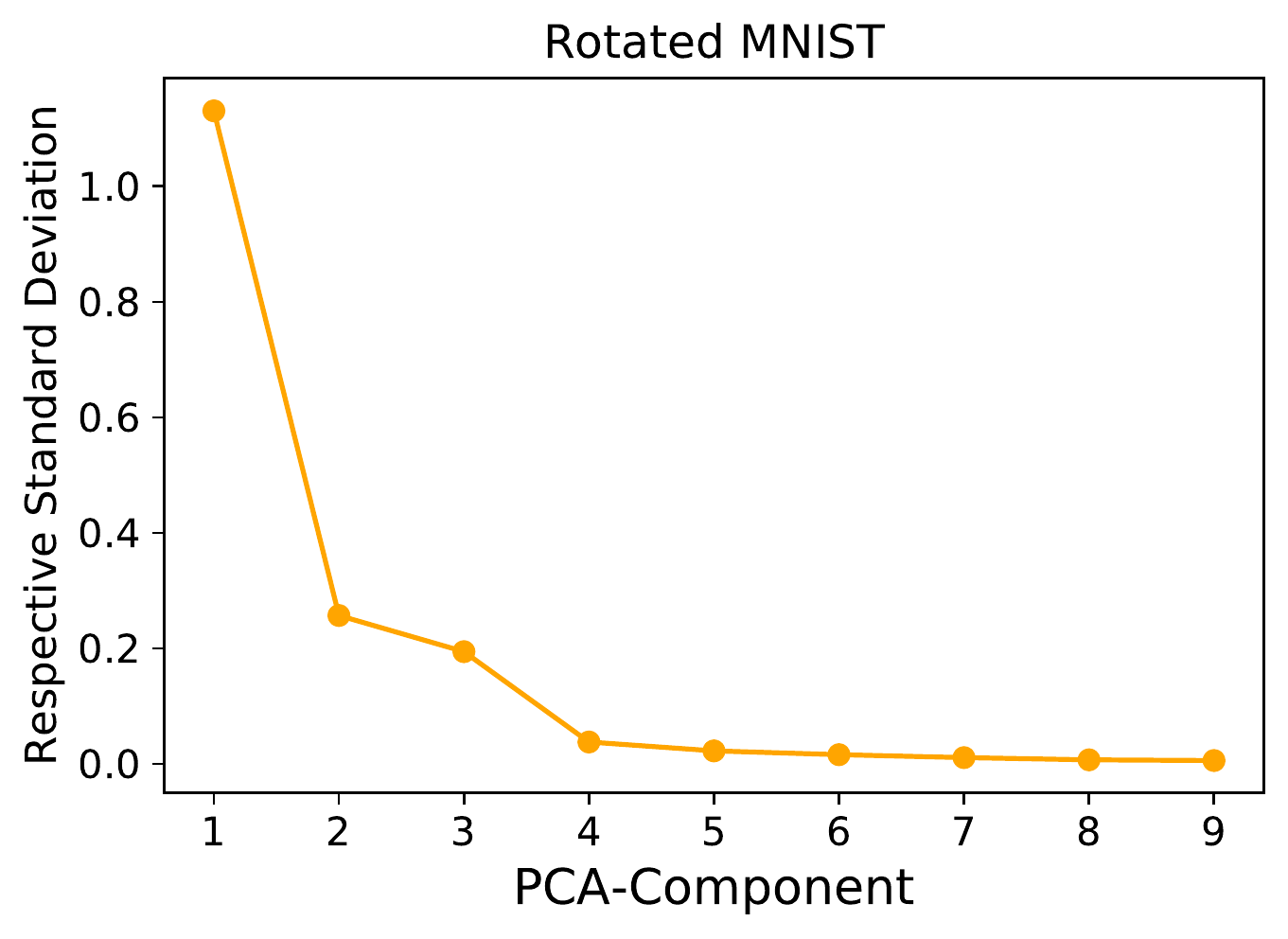}}
 \end{center}
 \vspace{-0.5cm}
\caption{{\bf Left:} Pointcloud of first three PCA components of our rotated MNIST dataset. Highlighted in orange are the orbits of multiple digits eight. Gray points correspond to the other digits present in this dataset. {\bf Right:} The standard deviation on the generators identified from this pointcloud for the digit eight.}
\label{fig:MNIST}
\end{figure}

Our analysis proceeds as follows: We consider a subset of the rotated MNIST dataset, consisting of $2000$ images of $8$ and their rotated versions $D_8.$ Note that such a subset of the dataset easily emerges when doing a classification task. On the entire rotated MNIST dataset $D_{\rm all}$ we perform PCA and consider the first three components. We apply this PCA transformation on the datasets containing only several rotated images of a single digit, e.g.~$D_8.$ A visualisation of the orbits associated to several digits eight can be seen in Figure~\ref{fig:MNIST}. On this pointcloud of digits eight, we now perform the remaining steps of our algorithms and find that the dominant generator is given by an $SO(2)$ rotation:
\begin{equation}
 G=\left(
\begin{array}{c c c}
-0.06& -0.00  & -0.07\\
 0.01& -0.01&  1.00  \\
 0.08& -0.99&  0.04 
\end{array}
\right).
\end{equation}
The respective standard deviations can be found in Figure~\ref{fig:MNIST} on the right. We clearly identify the generator of $SO(2)$ as the dominant generator.

\subsection{Discrete symmetries -- CICYs}
 
To conclude this section we briefly return to the example of CICYs discussed in Section~\ref{sec:cicys}. Per construction the symmetries acting are discrete rather than continuous. To identify underlying symmetries -- earlier referred to as identities (cf.~\eqref{eq:cicyidentities}) -- one needs to match identical transformations in different orbits acting on the input space. As our input dataset is precisely generated by these identities and such different representations are mapped to the same cluster in the embedding layer, our network does identify these identities. It will be interesting to analyse whether the network finds additional symmetries and identities which are yet unknown. However, this would require a different training approach with differently prepared datasets which we leave for future work.

\section{Conclusions}
\label{sec:conclusions}
Detecting symmetries in an automated fashion removes the necessity for domain knowledge associated to a particular data product. Such domain knowledge often might not be of existence or has been the outcome of scientific efforts such as in the development of the quark model~\cite{eightfold}. In this article we introduced a method on how to detect symmetries with only very limited domain knowledge. The required domain knowledge was to be able to perform a `simple' classification task which we think is often a realistic starting point.

We have discussed examples of basic symmetries appearing in physics such as rotational groups and $SU(2).$ The structure in the embedding layer does reveal these symmetries and hence provides orbits on the input space which are generated by these symmetries. In a second step we were able to pinpoint the nature of these continuous symmetries by our regression algorithm. Beyond rotational groups and $SU(2)$ we find that the embedding layer can be used to identify classes CICY-manifolds. It remains to be seen whether these methods can establish new identities in the case of the classification of $n$-folds which is unknown to this date. For this analysis, we introduced a novel graph representation for CICYs which removes several redundancies of the matrix representation used up to now. In passing we note that this provides the first application of graph neural networks in string theory. We have not yet explored the full potential on other ML work on this dataset with this representation (cf.~\cite{He:2017set,Ruehle:2017mzq,Bull:2018uow,Klaewer:2018sfl,Bull:2019cij,Brodie:2019dfx} for other ML applications on the CICY dataset).

Another observation which appeared in this analysis is that the neural network has found a way to calculate topological invariants as required by Wall's theorem which formalises how complex manifolds are completely characterised. We have not yet investigated this avenue but want to highlight that it will be exciting to compare these two complimentary approaches to classification. In which situations does a neural network obtain use such mathematically rigorous ways of classification?

We have seen that an important ingredient in our analysis are dimensional reduction tools -- here in particular TSNE~\cite{tsne}. It remains to be seen in the future which additional structures TSNE and other techniques can reveal on datasets in mathematical physics, similar to structures seen in autoencoders~\cite{Mutter:2018sra}. 

Putting this method into perspective, we can find that our results can be improved with augmenting the pointclouds. Additional points can be obtained if an equation generating these orbits is known. In this context it might be useful to utilise the techniques recently described in~\cite{wetzel2020discovering}. Furthermore, our technique of identifying symmetries is useful to determine which symmetry equivariant architecture (cf.~\cite{cohen2016group}) promises to be efficient for more sophisticated classification tasks. Beyond classification, another application in machine learning for utilising symmetries which has recently been proposed is in the context of reinforcement learning~\cite{pol2020plannable}. In either case, it promises to be extremely interesting to see which other symmetries can be found in every day and scientific datasets, going beyond a standard rotational invariance such as we discussed in the context of MNIST.

This is a proof of concept paper presenting several ways of identifying underlying symmetries in the data. Further scrutiny of these methods for other symmetries is in order. Now, it is even more tantalising to find out the underlying symmetry structures neural networks are dynamically using to achieve their remarkable performance.

\section*{Acknowledgments}
We would like to thank Per Berglund, Harold Erbin, Andre Lukas for useful discussions. Significant parts of this work were performed during the workshop the Data Science revolution at the Simons Center for Geometry and Physics in Stony Brook.

{\bf Note:} We are aware that Danilo Rezende and collaborators are working on similar questions related to identifying symmetries with neural networks.

\bibliography{NewBib} 

\bibliographystyle{JHEP}

\end{document}